\input harvmac.tex

\input epsf.tex
\def\figin{\epsfcheck\figin}\def\figins{\epsfcheck\figins}
\def\epsfcheck{\ifx\epsfbox\UnDeFiNeD
\message{(NO epsf.tex, FIGURES WILL BE IGNORED)}
\gdef\figin##1{\vskip2in}\gdef\figins##1{\hskip.5in}
\else\message{(FIGURES WILL BE INCLUDED)}%
\gdef\figin##1{##1}\gdef\figins##1{##1}\fi}
\def\DefWarn#1{}
\def\figinsert{\goodbreak\midinsert}
\def\ifig#1#2#3{\DefWarn#1\xdef#1{fig.~\the\figno}
\writedef{#1\leftbracket fig.\noexpand~\the\figno}%
\figinsert\figin{\centerline{#3}}\medskip\centerline{\vbox{\baselineskip12pt
\advance\hsize by -1truein\noindent\footnotefont{\bf Fig.~\the\figno:} #2}}
\bigskip\endinsert\global\advance\figno by1}



\def\p{\partial}
\def\mod{{\rm mod}}


\def\IL{\relax{\rm I\kern-.18em L}}
\def\IH{\relax{\rm I\kern-.18em H}}
\def\IR{\relax{\rm I\kern-.18em R}}

\def\CN {{\cal N}}

\def\CF {{\cal F}}
\def\CJ {{\cal J}}

\def\CL {{\cal L}}

\def\CO {{\cal O}}

\def\CH {{\cal H}}

\def\CA{{\cal A}}

\def\CW{{\cal W}}
\def\IZ{{\bf Z}}

\def\CN {{\cal N}}

\def\CO {{\cal O}}

\def\lieg{{\bf \underline{g}}}
\def\p{\partial}
\def\lieg{{\bf \underline{g}}}

\def\mod{{\rm mod}}

\def\vol{{\rm vol}}

\def\im{{\rm Im}}

\font\manual=manfnt \def\dbend{\lower3.5pt\hbox{\manual\char127}}

\def\half {{1\over 2}}
\def\sdtimes{\mathbin{\hbox{\hskip2pt\vrule height 4.1pt depth -.3pt
width
.25pt
\hskip-2pt$\times$}}}
\def\p{\partial}

\def\ra{\rangle}
\def\rara{\rangle\rangle}

\def\kbos{\kappa}
\def\ksusy{k}


%
\lref\affleck{I. Affleck and A. Ludwig, ``The Kondo Effect,
Conformal Field Theory, and Fusion Rules,''
Nucl.Phys.B352:849-862,1991 ; ``Critical Theory of Overscreened
Kondo Fixed Points,'' Nucl.Phys.B360:641-696,1991}

\lref\ars{A. Alekseev, A. Recknagel, and V. Schomerus,
``Non-commutative World-volume Geometries: Branes on SU(2) and
Fuzzy Spheres,''hep-th/9908040; ``Brane Dynamics in Background
Fluxes and Non-commutative Geometry,'' hep-th/0003187}

\lref\asii{A. Alekseev and V. Schomerus,
``D-branes in the WZW model,'' hep-th/9812193;
Phys. Rev. {\bf D60}(1999)061901}
\lref\as{A. Alekseev and V. Schomerus,
 ``RR charges of D2-branes in the WZW model,''
hep-th/0007096}
\lref\arsrev{
A. Yu. Alekseev, A. Recknagel, V. Schomerus,
``Open Strings and Non-commutative Geometry of Branes on Group Manifolds,''
hep-th/0104054 ; Mod.Phys.Lett. A16 (2001) 325-336}

\lref\emss{ S.~Elitzur, G.~Moore, A.~Schwimmer and N.~Seiberg,
``Remarks On The Canonical Quantization Of The
Chern-Simons-Witten Theory,'' Nucl.\ Phys.\ B {\bf 326}, 108
(1989). }

\lref\atiyahhirz{M.F. Atiyah and F. Hirzebruch, ``Vector Bundles
and Homogeneous Spaces,'' Proc. Symp. Pure Math. {\bf 3} (1961) 53.}

\lref\atiyahbook{M. Atiyah, {\it K-theory}, Addison-Wesley, 1989. }

\lref\atiyahsegal{M. Atiyah and G.B. Segal, unpublished.}

\lref\bds{C. Bachas, M. Douglas, and C. Schweigert,
``Flux Stabilization of D-branes,''hep-th/0003037 }
\lref\birke{L. Birke, J. Fuchs, and C. Schweigert,
``Symmetry breaking boundary conditions and WZW
orbifolds,'' Adv.\ Theor.\ Math.\ Phys.\ { 3}, 671 (1999);  hep-th/9905038}
\lref\bm{P. Bouwknegt and V. Mathai,  ``$D$-Branes, $B$ Fields
and Twisted $K$-Theory,'' JHEP {\bf 0003:007,2000}, hep-th/0002023.
}
\lref\bcmms{
P. Bouwknegt, A. L. Carey, V. Mathai, M. K. Murray, D. Stevenson,
``Twisted K-theory and K-theory of bundle gerbes,''
 hep-th/0106194 }
\lref\cartaneilenberg{H. Cartan and S. Eilenberg,
{\it Homological Algebra}, ch. XV  }

\lref\cftbook{P, Di Francesco, P. Mathieu,
D. S\'en\'egal, {\it Conformal Field
Theory} Springer, 1997
}

\lref\dmw{E. Diaconescu, G. Moore, and E. Witten, ``$E_8$ Gauge Theory,
and a Derivation of $K$-Theory from $M$-Theory,'' hep-th/0005090}

\lref\douglasnc{M. Douglas, ``D-Branes in Curved Space,'' hep-th/9703056;
``D-branes and Matrix Theory in Curved Space,''hep-th/9707228;
``Two Lectures on D-Geometry and Noncommutative Geometry,''
hep-th/9901146.}

\lref\duff{M. Duff et. al. }
\lref\epstein{Epstein and N. Steenrod, {\it Cohomology Operations},
Ann. Math. Stud. {\bf 50}, Princeton Univ. Press}

\lref\fffs{G. Felder, J. Fr\"ohlich, J. Fuchs, C. Schweigert,
``The geometry of WZW branes,''hep-th/9909030,
J.Geom.Phys. 34 (2000) 162-190}

\lref\fss{J. Fuchs, B. Schellekens, and C. Schweigert,
``From Dynkin diagram symmetries to fixed point
structures,'' hep-th/9506135}

\lref\fuchs{J. Fuchs, {\it Affine Lie Algebras and Quantum Groups},
Cambridge Univ. Press}

\lref\figeroa{J.M. Figeroa-O'Farrill and S. Stanciu, ``D-brane charge,
flux quantization and relative (co)homology,'' hep-th/0008038}
\lref\fredenhagen{S. Fredenhagen and V. Schomerus,
``Branes on Group Manifolds, Gluon Condensates, and twisted K-theory,''
hep-th/0012164}

\lref\freedwitten{D.S. Freed and E. Witten, ``Anomalies in String Theory
with $D$-branes,'' hep-th/9907189.}
\lref\freed{D. Freed,``Dirac Charge Quantization and Generalized
Differential Cohomology,''
  hep-th/0011220}
\lref\gawedzki{K. Gawedzki, ``Conformal field theory: a case study,''
hep-th/9904145}
\lref\gimon{O. Bergman, E. Gimon, and S. Sugimoto,
``Orientifolds, RR torsion, and K-theory,''
hep-th/0103183}
\lref\unwinding{R. Gregory, J. Harvey, and G. Moore, ``Unwinding
strings and T-duality of Kaluza-Klein and H-Monopoles,''
hep-th/970808; Adv.Theor.Math.Phys. 1 (1997) 283-297}
\lref\hm{J. Harvey and G. Moore, ``Noncommutative Tachyons and
K-Theory,'' hep-th/0009030.}
\lref\hopkins{M. Hopkins, Unpublished result. July, 2001}

\lref\kapustin{ A. Kapustin, ``$D$-branes in a Topologically
Nontrivial B-field,'' hep-th/9909089.}

\lref\mmspara{J. Maldacena, G. Moore, and N. Seiberg,
``Geometrical interpretation of D-branes in gauged
WZW models,'' hep-th/0105038}

\lref\mmsiii{J. Maldacena, G. Moore, and N. Seiberg, 
``D-brane charges in five-brane backgrounds,'' to appear}

\lref\mathaisinger{V. Mathai and I.M. Singer, ``Twisted
K-homology theory, twisted Ext-theory,'' hep-th/0012046}

\lref\marolf{D. Marolf, ``Chern-Simons terms
and the three notions of charge,''
hep-th/0006117}
\lref\myers{
R.~C.~Myers,
``Dielectric-branes,''
JHEP {\bf 9912}, 022 (1999)
[hep-th/9910053].
}
\lref\mm{R. Minasian and G. Moore,``K-Theory and Ramond-Ramond Charge,''
JHEP {\bf 9711}: 002, 1997; hep-th/9710230.}

\lref\morozov{A. Alekseev, A. Mironov, and A. Morozov,
``On B-independence of RR charges,'' hep-th/0005244}
\lref\polyakov{A.M. Polyakov, PLB59 (1975)82 ? or
Nucl. Phys. B120 (1977) 429 ? ITP DOESN'T HAVE THESE!!}
\lref\selfduality{G. Moore and E. Witten, ``Self-duality, RR fields
and $K$-Theory,'' hep-th/9912279.}

\lref\rosenberg{ J. Rosenberg, ``Homological Invariants of Extensions of
$C^*$-algebras,'' Proc. Symp. Pure Math {\bf 38} (1982) 35.}

\lref\stanciu{S. Stanciu, ``D-branes in group manifolds,''
hep-th/9909163}
\lref\standiuii{S. Stanciu, ``A note on D-branes in
group manifolds: flux quantization and D0 charge,''
hep-th/0006145}
\lref\taylor{W. Taylor, ``D2-branes in B fields,''
hep-th/0004141}
\lref\wendt{R. Wendt, ``Weyl's character formula for non-connected
Lie groups and orbital theory for twisted affine Lie algebras,''
math.RT/9909059}
\lref\wittenbaryon{E. Witten
``Baryons And Branes In Anti de Sitter Space,''
hep-th/9805112;JHEP 9807 (1998) 006}
\lref\wittenk{E. Witten, ``$D$-Branes And $K$-Theory,''
JHEP {\bf 9812}:019, 1998; hep-th/9810188.}
\lref\wittenstrings{E. Witten, ``Overview of K-theory applied to
strings,'' hep-th/0007175.}
%


%

\lref\dkps{ M.~R.~Douglas, D.~Kabat, P.~Pouliot and S.~H.~Shenker,
``D-branes and short distances in string theory,'' Nucl.\ Phys.\
B {\bf 485}, 85 (1997) [hep-th/9608024]. }

\lref\botti{R. Bott, ``A note on the Samelson product in the classical
groups,'' Comment. Math. Helv. {\bf 34} (1960). See Collected Works, vol. 1}
\lref\bottii{R. Bott, ``The space of loops on a Lie group,'' Mich.
Jour. Math. {\bf 5}(1958) 35}
\lref\smith{L. Smith, Topology {\bf 8} (1969)69;
Kane and Moreno, Can. J. Math. {\bf 6} (1988) 1331}

\lref\fqs{D. Friedan, Z. Qiu, and S.H. Shenker,
Phys. Lett. {\bf 151B}(1985)37}
\lref\knizhnik{P. di Vecchia, V.G. Knizhnik, J.L. Petersen, and
P. Rossi, Nucl. Phys. {\bf B253}(1985) 701}
\lref\kac{V.G. Kac and I.T. Todorov, Comm. Math. Phys.
{\bf 102}(1985)337}
\lref\itzhaki{
N.~Itzhaki, J.~M.~Maldacena, J.~Sonnenschein and S.~Yankielowicz,
``Supergravity and the large N
limit of theories with sixteen  supercharges,''
Phys.\ Rev.\ D {\bf 58}, 046004 (1998)
[hep-th/9802042].}

\lref\mn{J.~M.~Maldacena and C.~Nunez,
``Towards the large n limit of pure N = 1 super Yang Mills,''
Phys.\ Rev.\ Lett.\  {\bf 86}, 588 (2001)
[hep-th/0008001].
}

\lref\sen{A.~Sen, ``Tachyon condensation on the brane antibrane
system,'' JHEP {\bf 9808}, 012 (1998) [hep-th/9805170].}

 \lref\fs{J. Fuchs and C.
Schweigert, ``A classifying algebra for boundary conditions,''
hep-th/9708141, Phys.Lett. B414 (1997) 251-259 \semi
 ``Branes: from free fields to general
backgrounds,'' Nucl. Phys. { B530} (1998) 99,
hep-th/9712257\semi  ``Symmetry breaking boundaries I. General
theory,'' hep-th/9902132\semi ``Symmetry breaking boundaries. II:
More structures, examples,'' Nucl.\ Phys.\ B { 568} (2000) 543
[hep-th/9908025].}

\lref\superpotential{G.~Moore, G.~Peradze and N.~Saulina,
``Instabilities in
heterotic M-theory induced by open membrane  instantons,''
hep-th/0012104; E.~Lima, B.~Ovrut and J.~Park,
``Five-brane superpotentials in heterotic M-theory,''
hep-th/0102046.}

\lref\yokota{K. Abe and I. Yokota, ``Volumes of
compact symmetric spaces,'' Tokyo J. Math. {\bf 20} (1997) 87}

\lref\zhou{J.~Zhou,
``Page charge of D-branes
and its behavior in topologically nontrivial B-fields,''
hep-th/0105106.
}



\Title{\vbox{\baselineskip12pt
\hbox{hep-th/0108100}
\hbox{RUNHETC-2001-24 }
}}%
{\vbox{\centerline{D-Brane Instantons  }
\medskip
\centerline{and }
\medskip
\centerline{K-Theory Charges }
}}

\smallskip
\centerline{Juan Maldacena$^{1,2}$, Gregory Moore$^{3}$, Nathan
Seiberg$^{1}$ }
\medskip

\centerline{\it $^{1}$ School of Natural Sciences,}
\centerline{\it Institute for Advanced Study} \centerline{\it
Einstein Drive} \centerline{\it Princeton, New Jersey, 08540}

\centerline{\it $^{2}$ Department of Physics, Harvard University}
\centerline{\it Cambridge, MA 02138}

\centerline{\it $^{3}$ Department of Physics, Rutgers University}
\centerline{\it Piscataway, New Jersey, 08855-0849}

\bigskip
We discuss some physical issues related to the K-theoretic
classification of D-brane charges, putting an emphasis on the
role of D-brane instantons. The relation to   D-instantons
provides a physical interpretation to the mathematical algorithm
for computing K-theory known as the ``Atiyah-Hirzebruch spectral
sequence.'' Conjecturally, a formulation in terms of D-instantons
leads to a computationally useful formulation of K-homology in
general. As an application and illustration of this viewpoint we
discuss some issues connected with D-brane charges associated
with branes in WZW models. We discuss the case of $SU(3)$ in
detail, and comment on the general picture of branes in $SU(N)$,
based on a recent result of M. Hopkins.

\medskip

\Date{July 27,  2001}



\newsec{Introduction}

The present paper is intended to elucidate some physical
aspects of the K-theoretic formulation of D-brane charge.
In order to illustrate our general remarks we will apply
these considerations to the question of classification
of D-branes in WZW models. We summarize our results in the
present introduction.

\subsec{Formulating K-theory using D-instantons}

D-brane charge is classified by K-theory \refs{\mm,\wittenk}. Many
of the arguments for this are somewhat mathematical, and do not
emphasize the physical difference between K-theoretic, as opposed
to cohomological, classification of D-brane charges. One physical
interpretation has been given by Witten in terms of
brane-antibrane annihilation in \wittenk\ in the framework of the
Sen conjectures \sen. The present  note offers  a slightly
different (but related) viewpoint on the physical meaning of the
K-theoretic classification of D-brane charges. We will focus on
the following simple question: Suppose spacetime is a product $X
= \IR \times X_9$ where $X_9$ is a 9-dimensional space, possibly
noncompact. What are the possible cycles $\CW\subset X_9$ which
can be wrapped by a D-brane? Broadly speaking, the answer to this
question comes in two parts

\bigskip
\item{(A.)} The field theory on the D-brane must be
consistent. For example, it must be
anomaly free. This can put   restrictions on the possible
cycles on which ``free D-branes'' (i.e.
D-branes with no other branes ending on them)
can wrap.

\item{(B.)} We must identify branes which can be
dynamically transformed into one another.

\bigskip

In the classification of D-brane charges in type II string theory
the conditions (A) and (B) are implemented as follows:

\bigskip
\item{$(A)$} A D-brane can wrap $\CW\subset X_9$ only if
\eqn\conditap{ W_3(\CW) + [H]\vert_{\CW} =0 \qquad {\rm in }
\quad H^3(\CW,\IZ) } Here $W_3(\CW)$ is the integral
Stiefel-Whitney class\foot{The reader will be able to follow the
main points of this paper by ignoring this class (and ignoring
Steenrod squares). We include such torsion classes for
completeness. } of $T\CW$.  In particular, in the DeRham theory
$[H]_{DR}\vert_{\CW}=0$.  A similar condition applies in other
string theories. For example, in the bosonic string we must impose
condition \conditap\ without the $W_3(\CW)$ term.

\item{$(B)$} Branes wrapping homologically nontrivial $\CW$ can
nevertheless be unstable if, for some $\CW'\subset X_9$,
\eqn\conditbp{ PD(\CW \subset \CW') = W_3(\CW') +
 [H]\vert_{\CW'}.}
where the left hand side denotes the Poincare dual of $\CW$ in
$\CW'$.
In other string theories, like the bosonic string, the question
of stability is more complicated because they always include
tachyons.  It is not always simple to disentangle the instability
associated with the ordinary tachyon, which is always present, from
the instability of the brane.
\bigskip

(We remark parenthetically that in   the cohomological classification of
D-brane charges the principles $A,B$ are implemented
as follows. First, free  branes can wrap any homologically
nontrivial cycle in $X_9$. Second,  a brane wrapping a nontrivial cycle
is absolutely stable.
Thus, the homotopy classes of configurations
of  free branes can be labelled by
 $H_{cpt,*}(X_9; \IZ) \cong H^*(X_9;Z)$. )

In section two of this paper we will  justify conditions $(A)$
and $(B)$. Our point of view is not essentially new. Regarding
(A), the role of global anomalies has already been thoroughly
explained in \refs{\wittenk,\freedwitten,\freed}. Moreover, regarding
(B), the phenomenon of brane instability was already noted in
\dmw. The novelty, such as it is,  of the present note, is that
we give  a more precise description of the mechanism   of
K-theoretic brane instability. That mechanism is simply
instability  due to D-brane instanton effects, and the essential
phenomenon goes back to Witten's formulation of the baryon vertex
in the AdS/CFT correspondence \wittenbaryon.

In section three we go on to explain how conditions $(A)$ and
$(B)$, are related to K-theory. The essential point is that
imposing  conditions \conditap\ ``modulo'' conditions \conditbp\
is closely related to  a  mathematical algorithm for the
computation of $K$-theory known as the ``Atiyah-Hirzebruch
spectral sequence,'' (AHSS) including the AHSS  in the presence
of non-torsion $H$-fields. We therefore clarify the relation of
the AHSS to physics.

To be more precise, the rules $(A)$ modulo $(B)$ are related to
the computation of the AHSS at the third differential. It is
important to stress, however, that the AHSS only gives an
``approximation'' to K-theory since it is predicated on a
filtration and one must then solve an extension problem. Moreover,
while  the rules ``\conditap\ modulo \conditbp\  '' are related
to the cohomology of the third AHSS differential, they are on the
one hand stronger, and on the other hand, do not involve the use
of higher differentials. Conjecturally, a systematic
implementation of the rules $(A)$ modulo $(B)$, including the
possibility of branes within branes, is a complete formulation of
K-homology. The results of section 4 provide some evidence for
this conjecture.

Our improved understanding of the physical basis for the AHSS
clarifies  the physical basis for using the mathematical group
$K_H^*(X)$ for the classification of D-brane charges in the presence
of a cohomologically nontrivial $H$-field. While it is perhaps
well-accepted that $K_H^*(X)$ classifies D-brane charges when $H$
is nontorsion it is worth understanding the physical basis of
this claim more thoroughly. In his original proposal \wittenk\
Witten gave an argument which only applied to the case when $[H]$
is a torsion class. A more  direct argument was given by Kapustin
\kapustin, developing the ideas of \freedwitten. Unfortunately,
Kapustin's  method again only applies for $[H]$ torsion. It would
be extremely interesting to formulate a direct generalization of
his argument to the nontorsion case but this appears to be
nontrivial. In \bm\ P. Bouwknegt and V. Mathai pointed out that
the mathematical theory of $K_H^*(X)$ makes perfectly good sense
when $H$ is nontorsion and moreover has a natural formulation in
terms of $C^*$ algebras. (For further useful discussion along
these lines we recommend the paper of Mathai and Singer
\mathaisinger. See also \bcmms.) While these discussions fit in
well with our current understanding of D-branes, as discussed in
\refs{\wittenstrings,\hm}, they do not demonstrate the physical
relevance of $K_H^*(X)$.

By giving a clearer
physical foundation for the AHSS for nontorsion $H$-fields
we are giving further support to the general claim
that $K_H^*(X)$ is the correct D-brane charge group.
In fact,   our clarification of the physical
basis of $K^*_H(X)$ leads to some insights into
the {\it limitations} of this group as a group of
brane charges.
In a companion paper \mmsiii\ we will examine critically
the physical meaning of $K^*_H(X)$ in the larger
context of M-theory.

Finally, we
  would like to clarify our use of the term ``instanton.''   The
D-branes fall into superselection sectors which are labeled by
their charges.  The configurations which we refer to as
instantons represent transitions between different such sectors,
and therefore they identify them.  Consider a transition from a
brane configuration $A_i$ to a brane configuration $A_f$. It is
achieved by an interpolation $A(t)$ with the boundary conditions
$A(t_i)=A_i$ and $A(t_f)=A_f$.  We refer to the interpolation
$A(t)$ as an instanton.  It is also a D-brane in space $X_9$.

Our discussion depends only on the topology of $X_9$ and of the
various branes $A$ and $A(t)$ and not on their detailed
geometry.  If more information about the dynamics is given, then
the D-branes $A$ should be stationary solution of the equations of
motion, and $A(t)$ must satisfy the time dependent equations of
motion where the parameter $t$ is interpreted as the time. Then
an important distinction should be made between two situations
regarding the interpolation $A(t)$:
\item{1.}  When the transition between $A_i$ and $A_f$ is a
classically allowed transition the interpolation $A(t)$ should
satisfy the equations of motion with Lorentzian signature time.
\item{2.}  When the transition between $A_i$ and $A_f$ involves
tunneling and is not allowed classically $A(t)$ should satisfy
the equations of motion with Euclidean signature time $t$.

The traditional use of the term ``instanton'' is only for tunneling
transitions but we will use the term for every interpolation.

\subsec{Applications to D-branes in WZW models}

WZW models are the preeminant example of string backgrounds with
cohomologically nontrivial nontorsion $H$-flux. Moreover, they
are solvable conformal field theories. They are therefore very
natural examples in which to study ideas of the connection
between D-branes and K-theory. In section 4 we use the
description of allowed brane wrappings (``rule $A$ modulo rule
$B$'') to compute $K_H^*(SU(2))$ and $K_H^*(SU(3))$. While
$K_H^*(SU(2))$ is well-known, the result for $SU(3)$ is new. We
discuss in detail some of the necessary topology of $SU(3)$ in
order to establish the result.

We then proceed to interpret the $SU(3)$ classes in
terms of D-branes. In order to do this we review
(and slightly extend) the beautiful theory of
D-branes in group manifolds which has been developed
by several authors over the past few years. This we
do in sections five and six.

In section 7 we introduce a new set of D-branes in
the WZW theory which we call ``parafermionic branes.''
The discussion here is a programmatic description
of an extension of the ideas of \mmspara. Much work
remains to be done.

In section 8 we compare the CFT results on D-branes
with those of K-theory. In particular, we describe a
recent result of M. Hopkins computing $K_H^*(SU(N))$
for all $N$. The result of Hopkins fits in very
naturally and beautifully with the physical picture
of D-branes in WZW theory, although the detailed
CFT construction of a   set of D-branes producing
the full set of K-theory charges remains to be
done.

\newsec{Physical justification of conditions $(A)$ and $(B)$}

The condition $(A)$ is well-known. In the DeRham theory it simply
comes about since the $H$-field must be trivialized on the
D-brane by the equations of motion $H\vert_{\CW} = d(F+B)$, and
hence $[H]\vert_{\CW}=0$. At the level of integral cohomology the
condition \conditap\ follows from the cancellation of global
anomalies for fundamental open strings ending on the D-brane, as
shown in \refs{\freedwitten,\kapustin}.

To explain $(B)$, suppose there is a cycle $\CW'\subset X_9$ on
which 
\eqn\nonzero{ W_3(\CW') + [H]\vert_{\CW'} \not=0 .
}
As we have just
explained, we cannot wrap a D-brane on $\CW'$. Let us do so
anyway, with a D-brane {\it instanton }.

We can then cancel the global anomalies on the
D-brane worldvolume $\CW'$ by
adding a magnetic source for $F$ on $\CW\subset \CW'$
such that
\eqn\cancelanom{
PD(\CW \subset \CW') = W_3(\CW') + [H]\vert_{\CW'}.
}
The proof will be given presently. A D-brane ending
on $\CW$ provides just such a magnetic source.
Therefore, we
 arrive at the following picture:   A brane wrapping a
spatial cycle $\CW$ propagates in time,
and {\it terminates} on a D-instanton wrapping
$\CW'$. This means the brane wrapping the spatial cycle $\CW$
can be unstable, and decays due to  D-brane
instantons wrapping $\CW'$. This is illustrated in figure 1.

One should note that condition $(B)$ only means the brane
wrapping $\CW$ {\it might} decay. The brane wrapping $\CW$ could
be carrying other conserved charges that prevent decay to the
vacuum.
 This is related to the fact that we are only answering
the question of which cycles can be wrapped, and not addressing
the full question of which branes within branes are allowed. The
latter question is indeed answered by K-theory, and could be
analyzed by systematic use of the above viewpoint.

\ifig\figone{
A D-brane instanton wraps $\CW'$ which does not satisfy
\conditap.  Another D-brane world volume terminating on a cycle
$\CW\subset \CW'$  satisfying \conditbp\ is unstable (provided it
carries no other D-brane charges), and decays to the vacuum.
}
{\epsfxsize1.5in\epsfbox{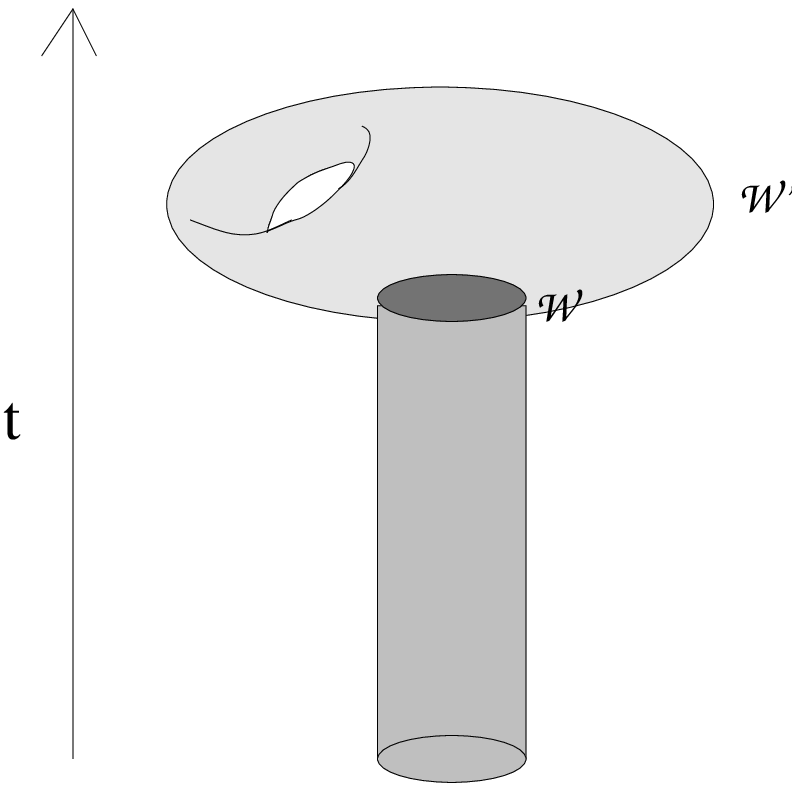}}

Let us now indicate why \cancelanom\ really does cancel the
anomalies. (See \freed\ for a more complete discussion.) In the
DeRham theory this is easy, since by definition of a magnetic
source $dF = 2\pi PD(\CW\subset \CW')$. To extend the argument to
the integral condition we proceed as follows\foot{The reader
uninterested in torsion should skip the remainder of this
section. }. For simplicity we put $H=0$, but allow $W_3(\CW')$ to
be nonzero.

Suppose a D-brane instanton wraps $\CW'$ with $W_3(\CW')$ nonzero.
Suppose another D-brane wraps $\CW \times R_+$ and terminates on $\CW'$.

We claim that in this situation we cancel the Freed-Witten anomaly
if $PD(\CW \to \CW') = W_3(\CW')$.
Recall that the Freed-Witten anomaly is based on the sign
ambiguity in the definition of the
of the path integral for worldsheet fermions. We thus
consider a family of open string worldsheets $S_t$, $0\leq t \leq 1$,
  ending on $\CW'$, where $S_0=S_1$. The boundaries of these
worldsheets $\p S_t$ sweep out  a 2-cycle $\Sigma_2=\p S  \subset \CW'$,
where $S$ denotes the total space of the family.
As shown in \freedwitten,
the holonomy of the  fermion determinants around $S^1$ is   a sign
\eqn\frwtt{
\exp[ i \pi \langle w_2(\CW'), [\Sigma_2] \rangle ] .
}
Freed and Witten then   extend the condition to require
that this factor be unity for any 2-cycle $\Sigma_2\subset \CW'$.
To proceed we will assume, again for simplicity, that
the anomaly is due to
a 2-cycle $\Sigma_2$ with $2 \Sigma_2 = \p D_3$ for some chain $D_3\subset\CW'$.
We can learn about $\Sigma_2$ by reviewing the definition of the
mod-two Bockstein map relating $w_2$ and $W_3$. Recall that we begin
by   lifting $w_2$ to an integral cochain $\bar w_2$ such
that
\eqn\frst{
\langle w_2, C_2 \rangle = \langle \bar w_2, C_2 \rangle \mod 2
}
for any 2-chain $C_2$.
Now $\langle \bar w_2, \Sigma_2 \rangle$ is   an {\it integer}. 
If $2C_2 = \p D_3$ we can now say:
\eqn\scnd{
2 \langle \bar w_2, C_2 \rangle =
\langle \delta \bar w_2, D_3 \rangle = 2 \langle W_3, D_3 \rangle
}
by definition of the Bockstein, and, this being an equation of integers,
we can divide by 2 to get
\eqn\thrd{
 \langle \bar w_2, C_2 \rangle =    \langle W_3, D_3 \rangle\mod 2
}
Now let us apply this to our cycle $\Sigma_2$. 
Note that $\langle W_3, D_3 \rangle $ is the oriented intersection
number of $D_3$ with $PD(W_3)$. So we see that an anomaly will
appear when there is a cycle $\Sigma_2$ that  links the
worldvolume $\CW$ of the terminating D-brane. In the presence of
a magnetic source there is an extra term in the path integral,
morally of the form $\exp[i \int_{\p S_t} A]$, where $S_t$ is the
open string worldsheet. This extra factor cancels the ambiguity
in the fermion determinant since 
\eqn\anomcan{
 \int_{\p D_3} F/2\pi = 1
\Rightarrow \int_{\Sigma_2} F/2\pi = 1/2 
}
and hence   the
magentic source along $\CW$ leads to a factor  $\exp[i
\int_{\Sigma_2} F ]=-1$, cancelling the ambiguous sign in the
fermion determinant.

\subsec{Examples}

In this section we illustrate the above
discussion with a few examples.

Consider  first the case of D0 brane number in a spacetime with
no torsion: The worldvolume of $N$ $D0$'s defines a 0-cycle $W_0$
which is  Poincare dual to $N x_9$ where $x_9$ generates
$H_9(X;Z)=Z$. \foot{We assume $X$ is connected. It is necessarily
oriented.} Naturally enough, we identify $N$ with the 0-brane
number in the approximation $H_{cpt}^{odd} \cong K^1$. However if
$\Sigma_3$ is any 3-cycle in spacetime on which $\int_{\Sigma_3}H
= k$, say, then we can consider a $D2$ instanton wrapping
$\Sigma_3$. This instanton will violate D0 charge since $k$
worldlines of $D0$ branes must end on $\Sigma_3$. The reason is
that these end on monopoles (=2+1 dimensional instantons) for the
$U(1)$ fieldstrength of the $D2$ brane so if $k$ worldlines of
$D0$'s end on a $D2$ at positions $x_i\in \Sigma_3$
 then $d F =2\pi \sum_{i=1}^k \delta^3(x-x_i)$.
In terms of cohomology we are identifying
\eqn\first{
N x_9 \sim N x_9 + H\wedge PD(\Sigma_3)
}
and since $\int_{\Sigma_3} H = k$ we know that
\eqn\second{
H\wedge PD(\Sigma_3) = k x_9
}
so that we have an identification $N \sim N+k$.
Consequently,   $D0$ brane number is only
defined modulo $k$, and the charge group has (at least)
$k$-torsion.

The same phenomenon shows that other D-instantons induce
$k$-torsion for other charges. For example, a $D2$ brane wrapping
$\sum_i n_i w_i$, where $w_i$ form a basis of $H_2(X,Z)$, has a
Poincare dual $7$-form $\omega_7 = n_i \omega_7^i$. Now we should
worry about $D4$ instantons wrapping various five cycles.  For
example, D4 instantons wrapping $\Sigma_5= (\sum_i a_i w_i) \times
\Sigma_3$ lead to the identification $n_i \sim n_i + k a_i$. The
reason is the same as before: A D4 instanton wrapping the
5-manifold $\Sigma_5$ violates the rule $[H]=0$. To account for
the trivialization $d\CF = dF + H $ we must have $D2$-brane
worldvolumes ending on the homology class $k (\sum a_i w_i)$ in
$\Sigma_5$.

As an example with   torsion in $H_{cpt}^*(X,\IZ)$,
and to make contact with \dmw\  let us consider a simple  example  with
$H=0$. Suppose that
$c = Sq^3(c_0)$  for some $c_0\in H^3(X,Z)$. Then $c$ is
$Sq^3$ closed, and hence corresponds to some K-theory
class. Nevertheless, it is also $Sq^3$-exact, and hence
its K-theory lift is zero. Since its K-theory charge
vanishes we expect a brane with
$PD(c)=\CW$ to decay. Physically, $PD(c)=Q_3$ is the
homology class of a spatial section of a D3-brane,
while $PD(c_0)=Q_6$ is the homology class of a cycle
which will be wrapped by a D5-brane  instanton.
Since $Sq^3(c_0)$ is nonzero, $Q_6$ is not $Spin^c$.
As we have explained, we  can cancel the
global anomalies by allowing the worldvolume $Q_3$
of the D3 brane to end on $Q_6$, provided
$PD(Q_3 \hookrightarrow Q_6) = W_3(Q_6)$.

\newsec{Relation of conditions $(A)$ and $(B)$ to K-theory}

In this section we will argue that conditions $(A)$ and $(B)$ are
in fact conditions of $K$-theory. We will show this by
demonstrating that $(A)$ and $(B)$ are refinements of the
procedure of taking $d_3$-cohomology in the AHSS for $K_H^*(X)$.

\subsec{Review: The mathematical Formulation of the AHSS}

Let us briefly review the AHSS. (For further background see \dmw.)

A $K$-theory class $x$ in $K^*(X)$ determines a system of
integral cohomology classes. If $x\in K^0(X)$ these are the Chern
classes $c_i(x)\in H^{2i}(X,\IZ)$. If $x\in K^1(X)$ there are
classes $\omega_{2i+1}\in H^{2i+1}(X,\IZ)$ related to
Chern-Simons invariants.

Conversely, given such a system of cohomology classes we may
ask whether it came from a K-theory class. The answer, in
general, is ``no.'' The AHSS is a successive approximation
scheme for computing  the necessary relations on the classes.
 This amounts to the following algorithm:

a.) In the first approximation
\eqn\firstappx{ \eqalign{ K^0(X) \sim   E^{\rm even}_1(X) &
:=\oplus_{j~ \rm even}  H^j(X,\IZ) \cr K^1(X) \sim  E^{\rm
odd}_1(X) & :=\oplus_{j~ \rm odd}  H^j(X,\IZ) \cr} }

b.) Then, for a certain differential\foot{The word
``differential'' means   $(d_3)^2=0$},  $d_3: H^j(X,Z) \to
H^{j+3}(X;Z)$,
 we compute
\eqn\ahss{
E^j_3(X):= \ker d_3\vert_{H^j}/\im d_3\vert_{H^{j-3}}.
}
and set $E_3^{\rm even} = \oplus_{j~ \rm even}E^j_3(X)$, etc. to
obtain the first correction:
\eqn\firstcorr{
 \eqalign{
K^0(X) \sim E^{\rm even}_3(X) &  := \bigl( {\rm Ker} ~d_3
\vert_{H^{\rm even} }\bigr) / \bigl({\rm Im} ~d_3\vert_{H^{\rm
odd}}\bigr)\cr K^1(X) \sim E^{\rm odd}_3(X)  & := \bigl( {\rm
Ker} ~d_3\vert_{H^{\rm odd} }\bigr) / \bigl({\rm Im}
~d_3\vert_{H^{\rm even}}\bigr)\cr } }

c.) Then, for a certain differential, $d_5: E_3^j(X,Z) \to
E_3^{j+5}(X;Z)$, we compute
\eqn\ahssi{
E^j_5(X):= \ker d_5\vert_{E^j_3}/\im d_5\vert_{E_3^{j-5}}
}
to obtain the next approximation, and so on.

d.) One keeps computing cohomology in this way to get
$E^j_{\infty}(X)$. (The procedure is guaranteed to stop after a
finite number of steps if $X$ is finite dimensional). The main
theorem states that  the ``associated graded group'' $Gr(K)$ is
given by
\eqn\mainthrm{\eqalign{
Gr(K_H^0(X)) &= \oplus_j E^{2j}_{\infty}(X)\cr
Gr(K_H^1(X)) & = \oplus_j E^{2j+1}_{\infty}(X)\cr}
}
We will explain the notation $Gr(K)$ presently. In  ``good
cases''  we can identify $Gr(K) = K$. See remarks below for the
meaning of ``good.'' The AHSS is useful because it is a clearly
defined computational algorithm, but it does have some drawbacks.

The first drawback is that one plainly needs to have a useful
expression for the differentials $d_3, d_5, \dots$. \foot{The
initial term of the spectral sequence is $C^*(X,h^*(pt))$, for
any generalized cohomology theory $h^*$.  Thus, following the
general procedure in \cartaneilenberg\ chapter XV,  one could in
principle extract the higher differentials. This appears to be
extremely difficult in practice.} A simple expression for the
differential $d_3$ is known. In  ordinary K-theory it was
identified in \atiyahhirz\ as $d_3 = Sq^3$. In \rosenberg\ it was
identified for twisted $K$-theory as $d_3 = Sq^3 + H $,  in the
context of $C^*$-algebra theory. This formula was rediscovered in
\refs{\dmw,\atiyahsegal}. Note that in the DeRham theory we obtain
the simple expression
\eqn\derhamdiff{
d_3(\omega) = [H]\wedge \omega
}
 Not much is known about the higher differentials in general.
There are scattered results for $H=0$, and it appears that
nothing further is known for $H$ nonzero. Fortunately, on
compact spin 10-folds at $H=0$ the higher differentials
are not needed \dmw.

A second drawback is that
in general the AHSS  only gives an ``approximation'' to
$K_H^*(X)$,  in the following sense.
Suppose we have a cell-decomposition, or simplicial
decomposition $X_0 \subset X_1 \subset \cdots \subset X_n$.
Then we define $K^*_{(p)}(X)$ to be the classes which
become trivial upon restriction to $X_p$. Obviously
$K^*_{(p+1)}(X)\subset K^*_{(p)}(X)$. The AHSS really
computes the ``associated graded space'' which is,
by definition
\eqn\agradeds{
\eqalign{
Gr(K^0_H(X)) &= \oplus_p K^0_{H,(p)}(X)/K^0_{H,(p+1)}(X) \cr
Gr(K^1_H(X)) &= \oplus_p K^1_{H,(p)}(X)/K^1_{H,(p+1)}(X) \cr}
}
In passing from \agradeds\ to the full $K$-theory group
one needs to solve an extension problem to obtain the
correct torsion subgroup. Fortunately, in many cases of
interest the extension problem is either not too severe or
even absent. However, there are important examples,
such as compact Lie groups of rank greater than two, 
and homogeneous spaces,  where
this complication can be significant. In the
physical context, the extension problem can
be quite important in the presence of orientifolds.
See  \gimon\ for a very interesting discussion of this point.

\subsec{Physical interpretation of the AHSS}

In this section we will explain the relation between taking $d_3$
cohomology in the AHSS and imposing the conditions $(A)$ modulo
$(B)$.

Recall we are working on spacetimes of the form $\IR\times X_9$.
Consider a brane wrapping a $p$-manifold $\CW\subset X_9$. One
may associate to $\CW$ several topological classes. First and
most obviously, $\CW$ has an associated homology cycle $Q(\CW)\in
H_{p}(X_9,Z)$. Since $X_9$ is oriented there is  a Poincare dual
integral cohomology class $\eta(\CW)\in H_{cpt}^{9-p}(X_9,Z)$.
Moreover, the brane wrapping $\CW$ has gauge fields and
consequently the D-brane charge is really associated with a class
in the $K$-theory of $X_9$. The homology class $Q$ can be
extracted from the $K$-theory class, since   $Q$ represents the
``support'' of the K-theory class. In physical terms, if the
D-brane is realized by tachyon condensation \sen, the support is
the locus where the tachyon field vanishes. We can then obtain a
cohomology class from $\eta=PD(Q)$. More generally, to a
collection $(\omega_1, \omega_3, \dots, \omega_9)$ we associated
a collection of branes by taking Poincar\'e duals to obtain a
collection of spatial cycles $(PD(\omega_1), \dots,
PD(\omega_9))$ around which $D8, D6, \dots, D0$-branes are
wrapped.

The interpretation of the kernel of $d_3$ is known from \dmw: A
necessary condition for a cohomology class $\omega\in
H_{cpt}^j(X_9;Z)$ to survive to $E^j_{\infty}$ is that
$d_3(\omega) = (Sq^3 + [H ]) \omega = 0 $. Clearly this is
implied by the anomaly cancellation condition \conditap. We
therefore interpret $d_3(\omega)=0$ as a (partial) requirement of
global anomaly cancellation. In fact, it is weaker than \conditap.

A natural question\foot{This question first came up in a
discussion with R. Dijkgraaf, July 1999.} now arises: How should
we interpret the {\it quotient } in \ahss\ by the image of $d_3$?
That is, what is the physics of the identification $\omega \sim
\omega + (Sq^3 + H\cup  ) \omega'$ in $E^j_{3}$ where $\omega'$
is a class of degree $(j-3)$ ? {}From the discussion in section 2
it is clear that the physical interpretation is that this
accounts for charge violation   due to $D$-instanton effects.
Again, dividing by the image of $d_3$ is weaker than the full
condition \conditbp.

\bigskip
{\bf Remark} We conclude with one remark. In the AHSS discussion
we need to choose a filtration of $K(X)$ (from a cell
decomposition) and we also need the higher differentials. On the
other hand,  in the physical discussion we didn't make any use of
filtrations, nor of higher differentials. It would be nice to
clarify the relation between these two procedures. Indeed, one
could probably deduce what the differential $d_5$ must be by
careful examination of conditions $(A)$ and $(B)$. In this
connection it is worth noting that $K$-theory is not easily
described by the cohomology of a natural functor from the
category of topological spaces to chain complexes.

\newsec{Application to WZW models:  Computing $K_H^*(SU(2))$ and
$K_H^*(SU(3))$}

Conformal field theories based on WZW models can be used to
construct vacua of type II string theory.  The simplest examples
involve a free time coordinate, a level $k$ WZW model of
$G=SU(2)$, $SU(3)$, or $SO(4)$, and other fields.  The restriction
to $SU(2)$, $SU(3)$ or $SO(4)$ comes from the requirement that the
total contribution to $\hat c$ from the spatial part of the
conformal field theory must be at most nine.  Other groups can be
used in a similar way provided their rank is less than nine and
their level $k$ is not too large.  Actually, a slight extension
of this framework allows us to use all possible groups with all
values of $k$.  The point is that if the dilaton varies linearly
in time, its contribution to $\hat c$ is less than one.  Allowing
such behavior of the dilaton, there is no restriction on $\hat c$
of the spatial part of the theory and all groups can be used.

Let us examine the D-branes in $SU(2)$ and $SU(3)$ in more detail.

\subsec{ $SU(2)$ }

In the level $k$ $SU(2)$ WZW model we have $[H] = k x_3$ where
$x_3$ is an integral generator of $H^3(SU(2);Z) = Z$. The
computation of $K_H^*(SU(2))$ from the AHSS is completely
straightforward and yields $Z/kZ$. (See \mathaisinger\ for the
computation in the $C^*$ algebra context.) This is easily
understood in terms of the general discussion of section 2:  The
D0 brane charge is violated by the D2 instantons.

This result is, of course, very closely related to results
discussed in the extensive literature on branes in WZW models
\refs{\affleck\asii\gawedzki\stanciu
\fffs\ars\bds\fs\as\birke-\fss}.  We will comment on some
relations to the literature in section 5  below.

The group $SU(2)$ is not sufficiently complicated topologically
to illustrate some important issues in the classification of
D-branes in WZW models. Some very interesting novelties already
arise when we try to extend the   discussion to the case of the
level $k$ $SU(3)$ WZW model. In the next two  sections we will
discuss the relation of D-branes to K-theory for the $SU(3)$
model.

\subsec{Branes in $SU(3)$: Some topological preliminaries}

In this section we summarize some topological facts which will be
useful in understanding the charge groups for D-branes in $SU(3)$
WZW models.

It is well-known that $SU(n)$ is a twisted product of odd spheres.
Rationally, it has the homotopy type of  $S^3\times S^5 \times
\cdots \times S^{2n-1}$.  Its integral cohomology is an exterior
algebra
\eqn\extalg{
H^*(SU(n);Z) = \Lambda_Z[x_3, x_5, \dots, x_{2n-1}]
}
where $\Lambda_Z[w_5, \dots, w_{2N-1}]$ is an exterior algebra on
generators $w_i$:
\eqn\exteriora{ \Lambda_Z[w_5, \dots, w_{2N-1}] = Z \oplus Z w_5
\oplus Z w_7 \oplus Z w_9 \oplus Z w_{11} \oplus Z w_5 w_7 \oplus
\cdots }
 and, in particular, has no torsion. (In DeRham theory the
 $x_i$ may be represented by suitably normalized
Maurer-Cartan forms.)
It might therefore come as a surprise that torsion in the
cohomology of $SU(n)$ (with other coefficient groups) is of
central importance in understanding the geometry of
WZW branes. The basic phenomenon can be understood in
fairly elementary terms as follows.

Let us find smooth submanifolds of $SU(3)$ representing
the Poincar\'e duals of $x_3$ and $x_5$.
The Poincar\'e dual to $x_5$ is easily written down:
it is an embedded $SU(2)$ subgroup of minimal
Dynkin index. For definiteness, we will take it to be
\eqn\sutwoinsuthree{
\pmatrix{ g& 0 \cr 0 & 1\cr} \qquad g\in SU(2)
}
This is a copy of $S^3$ and also represents a
 generator of $\pi_3(SU(3))$ and of $H_3(SU(3),Z)$.

The PD to $x_3$ is much more subtle.
A nice representative is given by the
submanifold $M_5\subset SU(3)$ of
{\it symmetric} $SU(3)$ matrices.
Let us make a few comments on this space.

First, $M_5$  intersects generic perturbations of the above
$SU(2)$ subgroup in a single point. For example, if  we take
 $$ \pmatrix{c & 0 & s \cr 0 & 1 &  0 \cr -s & 0 & c\cr}
 \pmatrix{ g & 0 \cr 0 & 1\cr} $$
as the representative of the generator of $H_3$, where $c^2 +
s^2=1$ and $s$ is   nonzero, and $g\in SU(2)$, then the
intersection with $M_5$ is readily seen to be the point
 $$\pmatrix{-c & 0 & s \cr  0  & -1 & 0   \cr    s & 0 & c \cr} $$
Since $M_5$ intersects a copy of $S^3$ in a single point we
conclude that it is a generator of $H_5(SU(3);Z)$.

Next, $M_5$ is homeomorphic to the homogeneous space
$SU(3)/SO(3)$ since (as one can show with a little bit of linear
algebra)  every symmetric $SU(3)$ matrix can be written in the
form $g g^{tr}$ for some $g\in SU(3)$. \foot{First prove the
analogous statement for $SU(2)$. Next, suppose $g$ is symmetric.
It can be diagonalized $g = u D u^{-1}$. We get $(u^{tr} u)D = D
(u^{tr} u)$. If the eigenvalues of $D$ are distinct $(u^{tr}
u)=d$ is diagonal and we take $g_0 = u d^{-1/2} D^{-1/2}$. If two
eigenvalues coincide we are reduced to the $SU(2)$ case already
considered. $\spadesuit$} Thus, $M_5$ is a submanifold of $SU(3)$
but can also be regarded as   a quotient  of $SU(3)$.

{}From the long exact sequence for fibrations we now get the exact
sequence
\eqn\hmtpyles{ \cdots \to \pi_i(SU(3)) \to \pi_i(M_5) \to
\pi_{i-1}(SO(3))\to \pi_{i-1}(SU(3)) \to \cdots }
This shows that
$M_5$ is simply connected, hence orientable, so
$H^0(M_5,Z)=H^5(M_5,Z)=\IZ$, while $H^4(M_5,Z)\cong H_1(M_5,Z)
=0$. This leaves unknown the middle cohomology groups. With a
little more work one can deduce $H^2,H^3$ from \hmtpyles\ but an
alternative argument, more useful for our purposes, can be given
as follows.

We begin by noting  that, over the reals,  one can represent
$x_3$ by the Maurer-Cartan form $x_3= {1\over 24\cdot \pi^2}
 {\Tr}_3(g^{-1} dg)^3$,
and this form {\it vanishes identically} on $M_5$ as a
differential form. To prove this note that
 $x_3$ is left invariant and
$M_5$ is  an homogeneous space. Evaluation at 1 involves
the trace of the commutator of two symmetric matrices
against a third symmetric matrix, and hence vanishes.
This fact suggests that $H^3(M_5,R)=0$, as indeed follows
from the Thom isomorphism theorem. Therefore $H_2(M_5,R)$ and
$H_3(M_5,R)$ are zero.  Nevertheless, $H_2(M_5,Z)$ is nonzero as
the
following simple argument demonstrates.
Consider
a cycle $\Sigma_2  $ in $M_5$  gotten by taking the intersection
\eqn\intersec{\Sigma_2 := SU(2) \cap M_5 }
where $SU(2)$ is the subgroup \sutwoinsuthree\ defined above.
One readily checks that $\Sigma_2$ is the set of matrices
\eqn\twosph{ \pmatrix{e^{i \phi} \cos\theta & i \sin \theta & 0
\cr i \sin \theta& e^{-i \phi} \cos \theta & 0 \cr 0 & 0 & 1 \cr}
}
and is hence a 2-sphere. This 2-sphere divides the $SU(2)$
subgroup of $SU(3)$ into hemispheres $D_+ \cup D_-$. By symmetry
$\int_{D_+} x_3 = \int_{D_-} x_3 = 1/2$. We may now conclude
that  $\Sigma_2$ defines a nontrivial homology class in $M_5$ .
For, if not,  we would fill it in with a 3-chain $\p
\CW_3'\subset M_5$ on which $x_3$ vanishes. But then $\Sigma_3=
D_+\cup \CW_3'$ would be a three cycle in $SU(3)$ with
$\int_{\Sigma_3} x_3=1/2$, but that is impossible for an integral
class. This suggests that $H_2(M_5,Z) = Z_2$, as is indeed the
case.

Using a little more technology (e.g.\ the Leray spectral
sequence) one obtains the full cohomology groups of $M_5$:
\eqn\cohogrp{
\eqalign{
H^0(M_5,Z) & = Z \cr
H^1(M_5,Z) & = 0 \cr
H^2(M_5,Z) & = 0 \cr
H^3(M_5,Z) & = Z_2 \cr
H^4(M_5,Z) & = 0 \cr
H^5(M_5,Z) & = Z \cr}
}
By the universal coefficient theorems
the cohomology  with mod two coefficients is
\eqn\cohogrpii{
\eqalign{
H^0(M_5,Z_2) & = Z_2 \cr
H^1(M_5,Z_2) & = 0 \cr
H^2(M_5,Z_2) & = Z_2 \cr
H^3(M_5,Z_2) & = Z_2 \cr
H^4(M_5,Z_2) & = 0 \cr
H^5(M_5,Z_2) & = Z_2 \cr}
}

Now we come to a crucial point:
The submanifold $M_5$ {\it is not $Spin^c$}!
We will now show this by establishing the
equivalent statement that a certain
characteristic class, $W_3(M_5)\in H^3(M_5;Z)$
is nonzero. In fact, we will show that
$W_3(M_5)=(x_3)\vert_{M_5} \not=0 $.
(The remainder of this section assumes
some knowledge of Steenrod squares. See
\dmw, section 4.1 for relevant background.)

While the integral
Steenrod operation  $Sq^3$ vanishes identically
acting on the integral cohomology of $SU(3)$,
the mod-two operation $Sq^2$ does not, and indeed
\eqn\steenrod{
Sq^2(r_2(x_3)) = r_2(x_{5})
}
where $r_2$ is reduction modulo two.
\foot{This  equation is related by
transgression to the equation $Sq^2(r_2(c_2)) = r_2(c_3)$
in the cohomology of $BSU(n)$, a relation which also played an
important role in \dmw. }
It follows at once that
 $Q=PD(x_3)$ is not $Spin$
since,  by definition,
$Sq^2(a) =  a \cup \pi^*(w_2(\CN))$
where $\pi: \CN \to Q$ is the normal bundle in
$SU(3)$.
Moreover,  $M_5$ is embedded in a spin (indeed, in a
parallelizable) manifold so we can identify
$w_i(\CN)=w_i(M_5)$.
Next,  note that $w_2(M_5)\in H^2(M_5;\IZ_2)$ is nonzero,
and yet $H^2(M_5;\IZ)=0$, so $w_2$ does not have
an integral lift and hence $W_3(M_5)$ is nonzero.
In order to see the relation of $x_3$ and $W_3$
we   combine the Wu formula
$Sq^2(r_2(x_3)) = w_2 \cup r_2(x_3)$ with
\steenrod, and use the fact that $x_5\vert_Q$ generates
$H^5$, to conclude that $(x_3)\vert_{M_5}$ is nonzero.
Now,
the Euler class of $Q=PD(x_3)$ is
represented by $x_3\vert_Q$, by Poincare duality.
On the other hand, the Euler class of a real rank
three oriented vector bundle is given by
$W_3(\CN)= W_3(M_5)$.

In conclusion $W_3(M_5)= x_3\vert_{M_5}$ is nonzero, and indeed
generates $H^3(M_5,Z)$.

\subsec{Branes in $SU(3)$: Atiyah-Hirzebruch spectral sequence}

Let us now describe what information on  $K_H^*(SU(3))$ for $H=k
x_3$ can be gleaned from the AHSS. \foot{GM would like to thank
I. Brunner, G. Segal, and V. Schomerus, for discussions on such
computations for $SU(n)$ groups, and their comparison to WZW
models during August 2000. Very recently M. Hopkins has 
 computed $K_H(SU(N))$. His result shows that the higher
differentials are all nonzero. We describe his result, and our
interpretation of it, in section 8 below.} Since $Sq^3$ vanishes
the  differential is $d_3(\omega) = k x_3\cup \omega =H\cup
\omega$. Therefore
\eqn\eethree{
E_3^j = (\ker~ d_3: H^j\to H^{j+3})/(\im ~  d_3: H^{j-3} \to H^j)
}
is given by
\eqn\eegroups{
\eqalign{
E_3^{j=3}  &=  x_3Z / kx_3 Z  \cong Z_k \cr
E_3^{j=8} &  = x_3 x_5 Z/ kx_3\wedge x_5 Z\cong Z_k \cr}
}
contributing to $K_H^1$ and $K_H^0$, respectively.

Now, for $SU(3)$ there could in principle be a further
differential $d_5$. All we know is that this is a group
homomorphism
\eqn\deefive{
d_5: E_3^3 \to E_3^8
}
and therefore of the form $d_5(\omega) = s \omega$ for some
integer $s$.

Then in the AHSS
$$ K^1_H = \oplus_{\rm j~ odd} E_\infty^j =
\oplus_{\rm j~ odd} E_5^j \cong \ker(d_5) $$ $$ K^0_H =
\oplus_{\rm j~ even} E_\infty^j = \oplus_{\rm j ~ even} E_5^j
\cong Z_k/(\im d_5) $$

There are no extension issues in passing from the associated
graded to the full K-theory  because the associated graded is
concentrated in a single cyclic group, for a given parity of the
grading. Therefore  we conclude
\eqn\ambiguous{
K^0_h = K^1_h = Z_{k/s}
}
for some $s$ which divides $k$. Since $d_5$ is unknown, $s$ is
unknown. In fact, the results below show that $d_5$ is nontrivial.

One can also approach the problem using the six-term
Meyer-Vietoris sequence
for $K$-theory, but the most obvious decompositions
lead only to  \ambiguous.

\subsec{Branes in $SU(3)$: The physical computation}

Let us now consider the computation from the physical
approach, as summarized by the two conditions (A) and
(B) of the introduction.

We begin with the computation
of $K^1_H$, which is a refinement of the odd cohomology.
The Poincare dual of $x_5$ is an $SU(2)$ subgroup,
and $H$ restricts to a nontorsion class. We cannot
wrap branes on this cycle. The Poincare dual of $x_3$
can be represented by the submanifold $M_5$ of
symmetric $SU(3)$ matrices, as discussed above.
Implementing condition (A), the cancellation of
global anomalies requires
\eqn\physcond{
[H]\vert_{\CW} + W_3(\CW) = 0
}
where $\CW $ is the brane worldvolume. As we have
stressed, $W_3(M_5)$ is not zero
so, if $H=kx_3$ then the condition \physcond\ becomes
\eqn\physcondii{
r( k W_3 + W_3) =0
}
where $r$ is
 the number of times the worldvolume is wrapped. Thus we find that
if $k$ is odd, there is no extra condition on $r$,
while if $k$ is even then $r=0$ modulo 2.
In the IIA theory in a spacetime with
$X_9= SU(3)\times \IR$ we should think of these as D6 branes. They
wrap the 5-cycle $\CW$ in $SU(3)$ and the line.

Now let us implement condition (B) from the introduction.
In this case D8 instantons wrapping $SU(3) \times R$ violate
D6 brane number by $k$ units, as discussed above.
We therefore conclude that the physical picture coincides with
K-theory if
\eqn\predicti{
K^1_{H=k x_3}(SU(3))  = \cases{
 \IZ/k\IZ & $k$ odd  \cr
  2\IZ/k\IZ & $k$   even \cr}
}

Let us now consider the computation of $K^0_H$, a refinement of the
even cohomology of $SU(3)$. The even degree cohomology groups of
$SU(3)$ are $H^0$ and $H^8$. The Poincare dual of $1 \in H^0$
is all of $SU(3)$. Again, $H$ restricts to a nontorsion
class and we cannot wrap branes on this cycle. Let us consider
$H^8$. These correspond to $D0$-branes in IIA or $D1$ branes
wrapping $\IR$ in $IIB$. In any case, there is no question
of global anomalies, but we must worry about the effects of
D-instantons. For definiteness we will focus on the IIA
interpretation.

First, we have the familiar story: D2 instantons wrapping the
$SU(2)$ subgroup violate $D0$ charge by $k$ units. However, when
$k$ is even there is another interesting instanton. As we have
stressed, $M_5$ is not $Spin^c$ so, if $k$ is even,  we cannot
wrap a D4 instanton on it. Let us do so anyway. Then there is a
global anomaly, due to $W_3(M_5)$. We can cancel this global
anomaly by adding a $D2$ instanton that ends on  a 2-cycle
$\Sigma_2\subset M_5$ whose homology class $Q_2=[\Sigma_2]$  is
Poincare dual (in $M_5$) to $W_3(M_5)$. Such a cycle is not a
boundary in $M_5$, but can be bounded  in $SU(3)$. Let us choose
such a boundary $\CW_3\subset SU(3)$ with $\p \CW_3=\Sigma_2$.
This chain is anomaly free and constitutes a {\it new type of
D-instanton that appears when $k$ is even. }

We now claim that the new D-instanton wrapping $\CW_3\cup M_5$
in fact  violates D0 charge
by $k/2$ units. This follows from the remarks around
\intersec. Recall that
the Maurer-Cartan form $x_3\sim {\Tr}(g^{-1} dg)^3$,
 vanishes identically on $M_5$ as a
differential form. Moreover,
the PD of $W_3(M_5)$ in $M_5$ can be represented by
the
 cycle $\Sigma_2 \cong S^2$ defined in \intersec.
This 2-sphere divides the $SU(2)$ subgroup of $SU(3)$ into
hemispheres $D_+ \cup D_-$ and we may choose $\CW_3 = D_+$
in the above discussion. But such a
 D2-instanton violates D0 charge
by $k/2$ units!
In this way we conclude that if K-theory is to match
physical expectations then we must have
\eqn\predictii{
K^0_{H}(SU(3)) = \cases{
   \IZ/k\IZ & $k$  odd \cr
   \IZ/{k\over 2}\IZ  & $k$   even \cr}
}

\newsec{Symmetry-Preserving branes in WZW models}

There is a large literature on  D-branes in WZW models. A 
partial reference list includes
\refs{\affleck\asii\gawedzki\stanciu
\fffs\ars\bds\fs\as\birke-\fss}. Here we briefly review and extend
the picture which has emerged, and comment on
  its relation to our discussion.

\subsec{Boundary states and conjugacy classes}

The WZW branes which are best understood are
those that can  be
described by boundary states $\vert B\rara\in \CH^{\rm closed}$
  satisfying the boundary conditions
\eqn\twinbc{ \biggl( J_{n} + \omega(\tilde J)_{-n} \biggr) \vert
B \rangle \rangle =0 \qquad n\in \IZ }
where $\omega$ is an automorphism of the affine Lie algebra.
Here  $J_n$ are the modes of the left-moving currents and $\tilde
J_n$ are the modes of the right-moving currents.  Boundary states
with $\omega=1$ are called ``symmetry-preserving branes.''
(However, any choice of $\omega$ leaves an unbroken affine Lie
algebra symmetry.)

In what follows we will
assume that we are working with the
diagonal modular invariant
\eqn\diagmodlr{ \CH^{\rm closed}= \oplus_{\lambda\in P^+_\kbos}
 V_\lambda\otimes \tilde V_{\lambda^*} }
Here the sum is over the irreducible
integrable representations $V_\lambda$ of the 
affine Lie algebra $\widehat{\lieg}_\kbos$ based on a simple 
Lie algebra $\lieg$. 
The notation $\tilde V$ indicates the right-moving component and $\kbos$ is the
level of the current algebra.

One constructs boundary states with good sewing properties
  using the Cardy theory. To begin, we construct the
character states, or ``Ishibashi states.'' Choosing $\omega=1$,
the character state $\vert \lambda\rara_I$ is labelled by
$\lambda\in P_\kbos^+$, and simply corresponds to the identity
operator under the isomorphism $V_\lambda\otimes  V_{\lambda}^*
\cong {\rm Hom}(V_\lambda, V_\lambda)$. The Cardy states are
expressed in terms of character states  as
\eqn\cardy{ \vert \lambda\rangle\rangle_C = \sum_{\nu\in P_\kbos^+}
{S_{\lambda\nu}\over \sqrt{S_{0\nu}}} \vert \nu \rangle\rangle_I }
where $S_{\lambda\nu}$ is the modular $S$-matrix and $\vert \nu
\rangle\rangle_I$ are the character states.

The  Cardy states  $\vert \lambda\rangle\rangle_C $
 have a beautiful geometrical interpretation, valid at large $\kbos$:
They correspond to branes wrapping certain regular conjugacy classes
 of the group familiar from the theory of the
Verlinde formula \refs{\asii,\gawedzki,\stanciu,\fffs}.
More precisely, these are the
regular conjugacy class $\CO_{\lambda,\kbos}$ of
the element $\exp[2\pi i (\lambda+\rho)\cdot H/(\kbos+h)]\in T$.
  For examples in level $\kbos$
$SU(2)$ theory the branes wrap conjugacy classes of
trace $ 2 \cos \chi $ with $\chi = \pi (2j+1)/(\kbos+2)$.
For $SU(N)$ we label representations by
Dynkin indices $\lambda = \sum_{i=1}^{N-1} a^i \lambda_i$,
with $\sum a^i \leq \kbos $. The conjugacy class
is the conjugacy class of
\eqn\explsutwo{ \exp {2\pi i  \over (\kbos+N) } (\lambda+\rho)\cdot H
= \exp\Biggl[ {2\pi i  \over (\kbos+N) } \sum_{l=1}^{N-1} (a^l +1)
\bigl(\epsilon_1 + \cdots + \epsilon_l -{l\over N}{\bf
1}\bigr)\Biggr] }
where $\epsilon_l = e_{ll}$ is the $l^{th}$ matrix unit along the
diagonal.

The above geometrical picture may be justified by a
straightforward generalization of the computation of appendix D
of \mmsiii , see also \fffs .
 We review the argument briefly. By the Peter-Weyl
theorem an orthonormal basis for $L^2(G)$ in the unit Haar
measure, for any compact group, is given by  the matrix elements
of unitary irreps:
 $\sqrt{d(\lambda)}D^\lambda_{\mu_L \mu_R}(g)$
where $d(\lambda)$ is the dimension of the representation
$\lambda$.  Therefore,
 a well-localized closed string state  is
\eqn\localized{
\vert g \ra = \sum_{\lambda\in P_\kbos^+,\mu_L, \mu_R}
 e^{-\epsilon c_2(\lambda)/\kbos} \sqrt{d(\lambda)}D^\lambda_{\mu_L
 \mu_R}(g)
\vert \lambda, \mu_L \mu_R\ra
}
This is to be regarded as a state in $\CH^{\rm closed}$ made
from vertex operators.  The exponential regularizing
factor is meant to suppress large representations,
which correspond to ``giant gravitons,''  rather than
well-localized closed string states.

The modular $S$-matrix satisfies
\eqn\modess{ \eqalign{ {S_{\lambda, \nu} \over S_{\lambda,0}} & =
\chi_{\lambda}(h_{\nu})  \cr }}
where, for compact, connected, simply-connected groups $G$,
$h_{\nu} = \exp[2\pi i {\nu+\rho\over \kbos + h} ]$. Therefore, we
have
\eqn\overlpa{ \langle g \vert \hat \lambda\rara_C =
\sum_{\lambda} e^{-\epsilon c_2(\lambda)/\kbos} \sqrt{d(\lambda)
S_{0\lambda}} \chi_\lambda^*(g)\chi_{\lambda}(h_{\hat \lambda}) }
Moreover, for $\kbos\to \infty$
\eqn\limdim{ {S_{\lambda, 0}\over S_{0 0}}  \rightarrow
d(\lambda) . } Hence branes are localized\foot{If we measure the
position of the brane with fundamental string states then we can
only determine the position to within a string length. Translated
into uncertainty in the coordinates in the maximal torus we find
$\delta \psi \geq {1\over\sqrt{\kbos}}$. One can also try to
measure the branes using D-branes themselves, i.e., using the
DKPS \dkps\ limit. See \ars\ for further discussion of this
point.}
 on the conjugacy classes $\CO_{\lambda,\kbos}$
 of the special
elements $h_\lambda$.

Another way to justify the geometrical picture has
been described by Felder et. al. in \fffs. Among other
things, these authors show  that the algebra of boundary operators
for boundary conditions $\vert \lambda\rara_C$
becomes the algebra of  functions on $G/T$ in the
large $\kbos$ limit.

Since $\CH^{\rm closed}$ is a representation of $G_L\times G_R$,
we may act on the boundary states to produce new states
$\rho_L(g_L)\rho_R(g_R)\vert \lambda\rara_C$. It is clear from
the above computation that, geometrically, these states
correspond to branes wrapping the rotated classes
$g_L \CO_{\lambda,\kbos} g_R$. They satisfy boundary conditions
of the type \twinbc\ where $\omega$ is an inner automorphism.

\subsec{Flux stabilization}

The above picture of D-branes raises a puzzle
first pointed out, and resolved, in \bds\ in
the case of $G=SU(2)$.
The conjugacy classes $\CO_{\lambda,\kbos}$
are homologically trivial in $G$. Since
D-branes carry tension one would expect the
branes to be unstable to shrinking. Nevertheless,
analysis of the open string
channel (in the superstring) shows that the branes are stable.

As pointed out in \bds, this paradox is resolved by   recalling
that   the brane can carry a topologically nontrivial line bundle.
In the present case there is a  Chan-Paton {\it line bundle} $L$
(i.e. the brane is {\it singly wrapped}) which  has $c_1(L ) =
\lambda + \rho$, where we identify $H^2(G/T;Z)$ with the weight
lattice. Therefore $[F/2\pi]$ is nontrivial, and consequently
terms of the form $\int_{\CO} F\wedge *F$ increase as the radii of
$\CO$ decrease. This increases the energy of the brane, and
balances the force due to the tension of the brane.

In appendix A we justify the above statements
in detail by showing  that the computation of \bds\
can be extended to higher rank groups.

\subsec{Blowing up D0 branes}


We will now review the ``blowing up phenomenon''
in the WZW model. We are reviewing here
the work of
\refs{\affleck,\ars, \bds,\as,\fredenhagen}.


Let us suppose the WZW model for $G$ a compact, connected, simply
connected Lie group is a factor in the conformal field theory of
a type II string compactification. Let us consider a system of
$N$ D0 branes in the group. It seems clear that the Cardy state
corresponding to the identity representation, $\lambda=0$, should
be   a D0 brane. We can see this from the shape computation, as
well as from the open string spectrum which is that of a single
D0 brane. It is reasonable to assign to this state D0 charge one.
By putting together $N$ of these states it is obvious that we can
have configurations  with charge $N$. What is not so obvious is
whether they can form a bound state. This problem was analyzed in
the  context of the Kondo model by Affleck and Ludwig \affleck .
What they found is the following. They started with $N$ D0 branes
and they considered giving a small non-commutative vacuum
expectation value to the D0 brane coordinates. The D0 brane
coordinates are $N\times N $ matrices. Affleck and Ludwig chose
these matrices to be $\epsilon S^a$ where $S^a$ is an $N\times N$
irreducible representation $\lambda$ of $G$. This amounts to
introducing the boundary interaction $\int dt \epsilon  S^a
J^a(t,\sigma =0)$ in the string worldsheet. This interaction is
marginally relevant \affleck  . At low energies the boundary
field theory flows to a new boundary fixed point and therefore a
new conformally invariant boundary condition. Affleck and Ludwig
argued that the resulting fixed point is the Cardy state
associated to $\lambda $. From an analysis of the DBI action one
can estimate the time for this decay to be of order $\Delta T
\sim \sqrt{\kbos}$ for natural initial conditions.

We summarize the Affleck-Ludwig argument here. Consider an open
string stretching between one D0 brane and $N$ D0 branes. The
boundary interaction described above results in a perturbation to
the Sugawara Hamiltonian
 $\Delta H = \epsilon  \sum_{n\in \IZ}
S^a J_n^a $ where $S^a$ represent the $N\times N$ generators of
$\lieg$ on $C^N$. Let us assume that this representation is
$V_\lambda\cong C^N$ with $\lambda\in P_+$. Affleck and Ludwig
observe that for the special value $\epsilon^* = { 2 \over \kbos+h}
$  the currents $\CJ_n^a:= J_n^a + S^a $ also form a  Kac-Moody
algebra which acts on  $(V_{\hat\lambda=0}\otimes C^N)$, where
$V_{\hat\lambda=0} $ is the $\hat \lambda =0$ representation of
the $J^a$ current algebra. This product space is isomorphic to a
highest weight representation with respect to the $\CJ_n^a$
current  algebra. Indeed, the representation of $\CJ_n^a$ is
isomorphic to $V_{\hat \lambda}$ where $\hat \lambda$ is the
image of $\lambda$ in the level $\kbos$ Weyl alcove under the affine
Weyl group. This last statement can be proved using the relation
between the fusion product and the cabling of Wilson lines in 3d
Chern-Simons theory\foot{One studies the Hilbert space $\CH$ of
the Chern-Simons theory on a disk with a source in a
representation $\lambda$ at a point $P$ in its interior \emss.
$\CH$ is determined as a product of $V_{\hat \lambda=0}$ on the
boundary of the disk and $V_\lambda$ at $P$, or as $\CH=V_{\hat
\lambda}$ at the boundary.}. This shows that the open string
spectrum for a string stretched between a single D0 brane and the
state resulting after the decay of $N$ D0 branes agrees with the
expected open string spectrum between a D0 brane Cardy state and
the Cardy state labeled by $\lambda$. Using a similar
construction one can show that one can reproduce the open string
spectrum between Cardy states labeled by $\lambda$ and $\lambda'$.

This argument has simpler incarnations in some limits. In the
large $\kbos$ limit and for a relatively small number of D0 branes
$1\ll N \ll \sqrt{\kbos} $ one can describe the process via the Myers
effect \myers , see \refs{\ars,\fredenhagen} for further
discussion. On the other hand, if $ 1 \ll \sqrt{\kbos} \ll N\ll \kbos$, we
can describe the resulting Cardy state in terms of a single
D-brane wrapping a conjugacy class with a suitable $U(1)$ line
bundle as in the flux-stabilization mechanism described above.

The Affleck-Ludwig argument is readily generalizable to
string field theory, this is most obvious in BSFT.

\subsec{Worldsheet supersymmetry}

It is important for our considerations that we are working with
the $\CN=1$ supersymmetric WZW model.  In this case a GSO
projection can make the D-branes stable.  Also, they become
orientable and a D-brane charge can be defined.

The Kondo model renormalization group flows generalizes readily
to the supersymmetric WZW  case as we now show. The
supersymmetric current algebra is \refs{\fqs,\knizhnik,\kac}:
\eqn\superwzw{ \eqalign{ I_a(z) I_b(w) & \sim {\ksusy
\delta_{ab}\over (z-w)^2} + i f_{ab}^{~~c} {I_c(w)\over z-w}
+\cdots \cr I_a(z) \psi_b(w) & \sim {i f_{ab}^{~~c}
\psi_c(w)\over z-w} + \cr \psi_a(z) \psi_b(w) & \sim {\ksusy
\delta_{ab}\over (z-w)} +\cdots\cr} }
Note $\{Q, \psi_a\} = I_a $ and $\{Q, I_a\}\sim \p \psi_a $. As is
well-known, one can define decoupled currents $J_a= I_a + {i\over
2\ksusy} f_{abc} \psi_b \psi_c$ which satisfy a current algebra with
level
\eqn\ksuper{ \kbos = \ksusy - h }
where $h$ is the dual Coxeter number.
The level that should be used in the K-theory analysis in section 4
is $\ksusy$.

In the Kondo model we use the supersymmetric perturbation $\int
d\tau S^a I_a(\sigma=0,\tau)$.  It is easy to think about the
flow as deforming the supercharge by $S^a\psi^a$.  At the
infrared fixed point we can apply the Affleck-Ludwig trick again
to $J_a \to J_a + S^a$.  Therefore, the fermionic part of the
boundary state is unchanged and the flow is effectively only in
the bosonic part of the state, although the fermions and the
bosons do not decouple along the renormalization group trajectory.

\subsec{Assigning D0 charge to the symmetry-preserving branes}

We now consider the problem of assigning a K-theory class to the
explicit conformal field theory boundary states discussed in
the previous subsections.

There are two ways to associate a D0 charge to a collection of
D0-branes in the level $k$ WZW theory. First, one can simply
compute the integral $\int_{\CO_{\lambda,k}} e^{F+B}$ appearing
the the standard WZ coupling of D-branes. This leads to quantum
dimensions, a puzzling result, which we will not try to address in
this paper.

A more straightforward approach is simply to invoke the
blowing-up phenomenon described in section 5.3. It is natural to
try to define  the D0 charge of the Cardy state labeled by
$\lambda$  to be the dimension of the representation $\lambda$, 
denoted $d(\lambda)$. 
Note, however, that at fixed $k$ there are only a finite number
of integrable representations. We therefore assume that the D0
charge is torsion and is   given by $d(\lambda) \mod n(\kbos,G)$
where $n(\kbos,G)$ is an integer depending on $\kbos$ and $G$. It
is clear from the existence of D2 instantons wrapping the $SU(2)$
subgroup of $G$ that the D0 charge will be $k$-torsion. (Recall
that $\kbos$ and $\ksusy$ are related as in \ksuper.) However, as
we saw in section 4, there can be more subtle instantons arising
from branes wrapping higher dimensional homologically nontrivial
cycles which nevertheless violate D0 charge. We will now describe
the result  for $n(\kbos,SU(N))$.

To begin, let us consider  the case of $G=SU(2)$. The classes
$\CO_{\lambda,\kbos}$ correspond to $g= \exp[i \chi \hat n \cdot
\vec \sigma]$ for $\chi = \pi l/(\kbos+2)$ where  $l=1,  \dots,
\kbos+1$. Together with $l=0$, corresponding to no D0 branes at
all, this suggests a charge group $Z_{\kbos+2}= Z_{\ksusy}$.

Unfortunately, this simple argument does not generalize to
higher rank groups.   A formula for $n(\kbos,SU(N))$ was
proposed by
 Fredenhagen and Schomerus \fredenhagen\ based on the
renormalization group flows in the
Kondo model \affleck, as reviewed above.  We will now offer an
alternative derivation. Our argument is also based on
the  validity of the
blowing-up hypothesis. Our argument is very  similar to an
argument given in \as.

We will use the  left $\times $ right rotational $G\times G$
symmetry mentioned at the end of section 5.1.  In particular, we
consider the simple action of rotating the conjugacy class by the
left action of an element of the center $z\in Z(G)$.

Now, the center of $G$ may be identified with a subgroup of the
group of automorphisms of the extended Dynkin diagram of $G$
\refs{\cftbook,\fuchs}. The center of $G$ rotates orbits, while
the automorphisms of the extended Dynkin diagram act on the
integrable level $\kbos$ representations.  The two actions are
related by $z \CO_{\lambda,\kbos} = \CO_{\lambda',\kbos}$. In
particular, for   $SU(N)$ one easily checks the identity
\eqn\rotatecc{
\exp\bigl[ {2\pi i \over \kbos+N} (\lambda' + \rho)\cdot H\bigr]
= z \sigma \exp\bigl[
{2\pi i \over \kbos+N} (\lambda + \rho)\cdot H\bigr]\sigma^{-1}
}
Here $z=e^{-{2 \pi i \over N}} 1_N$  generates $Z(SU(N))$, while
  $\sigma$ is a permutation matrix
taking $\sigma \epsilon_i \sigma^{-1} =  \epsilon_{i+1}$
(corresponding to the Coxeter element of the Weyl group).  The
transformation $\lambda \to  \lambda'$ acts on Dynkin indices as
\eqn\automorph{
\lambda = \sum_{i=1}^{N-1} a^i \lambda_i
\to (\kbos-\sum_{i=1}^{N-1} a^i)\lambda_1
+ a^1 \lambda_2 + \cdots + a^{N-2} \lambda_{N-1}.
}
This is the standard action of the $Z_N$ outer automorphism of
$\widehat{SU(N)}$ on weight vectors and it generates the spectral
flow transformations.

Now, if one considers the action of $z\in Z(G)$ on the Cardy
states, one finds that $z \vert \lambda\rara_C$ corresponds to
the rotated conjugacy class $z \CO_{\lambda,\kbos}$, as well as to
the state $\vert \lambda' \rara_C$. To prove this we use the
property of the modular $S$-matrix  \refs{\cftbook,\fuchs}
\eqn\modesssym{
S_{\lambda', \mu} = S_{\lambda, \mu} e^{-2\pi i (\lambda_1,\mu)}
}
where $\lambda_1$ is the fundamental representation.
The phase on the right hand side of
\modesssym\  is precisely
the action of rotation of left-movers
by $e^{-{2 \pi i \over N}}$ when acting
on the Ishibashi state
$\vert \mu \rangle\rangle_I$.

Now, a rigid rotation of a brane cannot change its D0 charge.
Therefore, we   require
\eqn\specflowi{
d(\lambda' ) = (-1)^{N-1}  d(\lambda) \mod n(\kbos,G) .
}
The origin of the relative sign is the following:
The orientation of a brane wrapping a
conjugacy class rotated from $\CO_{\lambda,\kbos}$ by
$z$ and a brane wrapping a conjugacy class
$\CO_{\lambda',\kbos}$ resulting from blowing up D0 branes
might differ.
Since the Coxeter element is a product of $(-1)^{N-1}$
reflections the relative orientation beteen  the
rotated  and the blown-up brane differs by
this factor.

It is not difficult to show that if two generic conjugacy
classes in $G=SU(N)$ are related by rotation $\CO = \CO' g$
then $g\in Z(G)$, and hence \specflowi\ (and its
spectral flow descendents) are the only conditions we
should impose.

The equation \specflowi\ is   a nontrivial constraint on
$n(\kbos,G)$. For $SU(2)$ we have $d(j) = 2j+1$ so

\eqn\specflowii{
2({\kbos\over 2}-j) +1 = - (2j+1) \mod n(\kbos)
}
leading to $n(\kbos) = \kbos+2 = \ksusy$. For $SU(3)$ we have
$d(\lambda) = \half (a+1)(b+1)(a+b+2)$ for
$\lambda = (a,b)$, while
and $\lambda' = (\kbos-a-b,a)$. Then
$d(\lambda) = d(\lambda')\mod ~n(\kbos)$
determines $n(\kbos)=\kbos+3=\ksusy $ for $\kbos$
 even and $n(\kbos) = (\kbos+3)/2 = \ksusy/2$
for $\kbos$ odd. For general $N$ we reproduce the
answer of Fredenhagen and Schomerus:
\eqn\fredschom{ n(\kbos, SU(N)) = {\rm gcd} \{ a_i := d(\kbos \lambda_1 +
\lambda_i)\} = {\kbos+N\over {\rm gcd} (\kbos+N, {\rm lcm}(1,2,\dots,
N-1)) } }

\bigskip
\noindent
{\bf Remarks}:

\item{1.}  The detailed connection to the
argument in \fredenhagen\ is the following.
They consider the representation $\kbos\lambda_1$.
This has $SU(N)$ fusion rules
\eqn\frone{ \kbos\lambda_1 \times \lambda_i = (\kbos\lambda_1+ \lambda_i)
\oplus ((\kbos-1)\lambda_1 + \lambda_{i+1} ) }
and $\widehat{SU(N)}_\kbos$ fusion rules
\eqn\fronep{
\kbos\lambda_1 \times \lambda_i =   ((\kbos-1)\lambda_1 + \lambda_{i+1} )
}
If one identifies the  ``charge group'' as $\IZ_x$ where $x$ is
the smallest integer such that
 \eqn\fredcrit{ d(\kbos\lambda_1)
d(\lambda_i)= d(((\kbos-1)\lambda_1 + \lambda_{i+1} )) \mod x }
then
one finds \fredenhagen, $x= {\rm gcd}(a_i)$, with 
$a_i = d(\kbos \lambda_1 + \lambda_i)$. To connect to our
argument one proves  the identity
\eqn\frtwo{
\sum_{i=1}^{N-1} (-1)^i a_i + (-1)^{N} d(\kbos \lambda_1) = -1
}
by using Weyl's product formula. It thus follows that
$d(\kbos\lambda_1) = (-1)^{N-1} \mod x$, so \fredcrit\ becomes
\specflowi.

\item{2.} It is important to realize that in a
charge group not all charges need be realized, and
some charges might be realized more than once.
The reader need only compute the charges
$d(\lambda) \mod ~n(\kbos,SU(3))$ for $\kbos=2$
to see some examples.

\newsec{WZW branes from outer automorphisms}

The branes defined by the boundary condition \twinbc\ have been
investigated in \refs{\fffs,\fredenhagen}. These branes wrap
twinned conjugacy classes associated with the automorphism
$\omega$. Here we comment on these branes, mainly to facilitate
comparison with our computation of $K_H^1(SU(3))$ in section 8.

\subsec{The shape of the symmetry-preserving boundary
states for outer automorphisms}

Let us  focus on the
case of  the outer automorphism $\omega_c$ of
$SU(3)$ given by complex conjugation.
By \twinbc\ the  representation of the zeromode algebra
 must be such that $\lambda_L = \lambda_R$.
Since we are working with   the diagonal modular
invariant theory, the weights must also be related by
$\lambda_R = \lambda_L^*$ to exist in the closed string
spectrum. Therefore the allowed representations
$\lambda$ are the symmetric weights
 with Dynkin labels $(n,n)$, $n=0,1,\dots, [\kbos/2]$.

According to \refs{\birke,\fffs} there is a natural basis of
solutions to \twinbc\ given by ``twinned character states'' (or
Ishibashi states) $\vert \lambda \rangle\rangle_I^{\omega_c} $
which are defined such that
\eqn\ihsibover{ \eqalign{ {}_I^\omega\langle\langle (b,b)\vert
q_c^{\half (H+\tilde H)} \vert (a,a) \rangle\rangle_I^\omega &
=\delta_{a,b} \chi_{(b,b)}(q_c)\cr {}_I\langle\langle (b,b)\vert
q_c^{\half (H+\tilde H)} \vert (a,a) \rangle\rangle_I^\omega & =
\delta_{a,b} \chi_{(b,b)}^\omega(q_c)\cr} }
where $\chi^\omega$ are the ``twinned characters'' of \fss. For
explicit expressions, see \refs{\fss,\birke}, and references
therein.

The corresponding ``Cardy'' boundary states with good
sewing properties are obtained using a modular transformation
matrix $S_{ab}^\omega$ (for $\lambda = (a,a), \mu = (b,b)$) given by
\eqn\twinnmod{ S_{ab}^\omega = {2\over \sqrt{\kbos+3}} \sin\bigl[
2\pi {(a+1)(b+1)\over (\kbos+3)}\bigr] }
(For $\kbos$ odd this is the S-matrix of an SU(2) theory at level
$[\kbos/2]$.) Explicitly,
\eqn\twinncard{
\vert (a,a)\rangle\rangle_C^\omega = \sum_{b}
{S^\omega_{ab}\over \sqrt{S_{0, (b,b)}}} \vert (b,b)\rara^\omega_I
}

We can now measure the shape of the branes $\vert
(a,a)\rangle\rangle_C^\omega$ by taking the overlap with a
well-localized closed string state.
 We parametrize elements
of $SU(3)$ as
\eqn\twinned{ g=
h \pmatrix{ \cos\theta & \sin \theta & 0 \cr
- \sin \theta& \cos \theta & 0 \cr
0 & 0 & 1 \cr} h^{tr}
}
then we have the overlap
\eqn\goverishi{ \langle g\vert (a,a)\rara_I^{\omega_c} = {\sin
2\phi(a+1)\over \sin 2\phi} }
where we have
defined an angle
\eqn\relateangl{
1 + 2 \cos \phi = - \cos \theta
\quad \Rightarrow \quad
\cos \phi = - (\cos \half \theta)^2
}
The necessity for this change of angles
lies in the difference between a diagram
automorphism and the complex conjugation automorphism.

At large $\kbos$ we find
\eqn\twinnoverlp{ \eqalign{ \langle g \vert (\hat a , \hat a)
\rangle\rangle_C^{\omega} & \sim {\kbos^3 \over \sin 2\phi}
\sum_{b=0}^{[\kbos/2]}    \sin((b + 1)\phi) \sin ((b+1)\hat \psi_a)
\cr & \sim {\kbos^3\over \sin 2 \phi}  \delta_{\rm periodic}(2\phi -
\hat \psi_a)\cr} }
where $\hat \psi_a = 2\pi (\hat a+1)/(\kbos+3)$
and we  have dropped overall numerical factors.

Thus, geometrically, the branes \twinncard\ are branes
wrapping the twinned conjugacy classes
\eqn\twinnclss{ \CO^\omega(t):=\{  h t h^{-1,*}: h \in SU(3)  \}
= \{  h t h^{tr}: h \in SU(3)  \} }
for $t\in T_2^\omega$, where
we have the maximal torus $$ T_2^\omega =  \{ \pmatrix{
\cos\theta & \sin \theta & 0 \cr - \sin \theta& \cos \theta & 0
\cr 0 & 0 & 1 \cr} \} $$

Note that the manifold $M_5$ of section 4.2 is just the twinned
conjugacy class of the identity. From \twinnoverlp\ we see that
the branes corresponding to the boundary states
$\vert (a,a)\rara_C^{\omega_c}$
wrap twinned conjugacy
  classes   localized at discrete values of $\theta$, whose values
can be read off from \twinnoverlp\ and \relateangl. For generic
$\theta$ the twinned conjugacy  classes   may be identified with
the
 7-dimensional  homogeneous
spaces of the form
  $SU(3)/U(1)$. They    are     $S^2$ bundles over the
twinned conjugacy class of the identity, $M_5$.
\foot{It can be useful to view these as ordinary conjugacy classes
in the disconnected component of $\IZ_2 \sdtimes SU(N)$.
See \wendt\ for several relevant results on these classes.}

\subsec{Spectrum, wrapping number, and gauge bundle
on the twinned branes}

The cohomology groups in \cohogrp\ show that
 we cannot put topologically 
nontrivial Chan-Paton line bundles on
$M_5$, hence we do not expect any further superselection sectors
for branes in $SU(3)$.

Using the above boundary states and
passing to the open string channel
we can examine the spectrum of the ``twinned brane.''
Unitarity of $S_{ab}^\omega$ shows that the unit operator
appears once.

Let us consider the masses and wrapping numbers of the
twinned branes. From conformal field theory
we can compute the mass of the twinned brane
relative to a D0 brane:
\eqn\twinnmass{ {S^\omega_{a0}\over S_{00}} = \sqrt{3}
{(\kbos+3)^{7/2} \over (2\pi)^3} \sin (2\pi {a+1\over \kbos+3}) }
which behaves correctly as a function of $\kbos$ in the sense that
for $a\ll \kbos$ we have a 5-dimensional object but for
$a \sim \kbos$ we have a 7-dimensional object.

On the other hand, using equation B.5 of the appendix
we see that an $L$-times wrapped brane on $M_5$ has
a mass
\eqn\expcted{ L \cdot   ({\kbos \over 2})^{5/2} 4 \sqrt{3\over
2}(2\pi)^3 {1\over (2\pi)^5} = {L\over 2} \kbos^{5/2} \sqrt{3}
{1\over (2\pi)^2} }
where the last factor in the right hand side is the D-brane tension.
We thus conclude that, for $a\ll \kbos$ the wrapping number is
\eqn\wrappi{
L=2a +2,
}
and is always even! On the other hand, for $a\approx \kbos/2 $ we must
rewrite $\sin (2\pi {a+1\over \kbos+3})= \sin (\pi {\kbos-2a+1\over
\kbos+3})$
from which we conclude
\eqn\wrappii{ L = \kbos-2a +1 \qquad  a\approx \kbos/2 }
If $\kbos$ is even then we can set $a=\kbos/2 - n, n=0,1,2$
and the wrapping number is $2n+1$, and always odd.
If $\kbos$ is odd then $a={\kbos-1\over 2} - n$ and the wrapping
number is $L=2n+2$ and always even.

Let us now compare these results on wrapping number with the
considerations of section 4.  We start with the bosonic string.
Here a necessary condition for being able to wrap branes is
equation \conditap\ without the $W_3(\CW)$ term
\refs{\freedwitten,\kapustin}. In our case $[H]|_\CW=0$ in
$H^3(\CW,Z)$ translates to $\kbos L\in 2Z$. Therefore, for $\kbos$ even
all values of $L$ are allowed, while for $\kbos$ odd only even values
of $L$ are consistent.  This is exactly the result found above in
the conformal field theory.

Turning to the superstring, we must recall from \ksuper\ that
$\ksusy = \kbos +3$
 determines the cohomology class of $H=dB$. This shows that
the parity of $\kbos$ (used here) is opposite to $\ksusy$. Bearing this
in mind and adding back the $W_3(\CW)$ term, which is not present
in the bosonic string, we again find the condition $\kbos L\in 2Z$.
Again, we see a perfect match between results of the conformal field
theory and the implications of the topology discussed in section
4.

The K theory considerations are based on anomaly cancellation and
the requirement that $\exp[i\int_{\Sigma} B]$ must be well
defined for worldsheets ending on the brane. Even though the
anomaly can only be seen at the level of {\it torsion} cohomology
classes, the boundary state ``knows'' about it. We find this
somewhat remarkable.

\newsec{Parafermionic branes}

Boundary conditions of the type \twinbc\ construct
rather special branes, preserving a full
current algebra symmetry. There are many other
branes in WZW theory one can construct by generalizing
the methods of \mmspara. (See also the works of
Fuchs and Schweigert et. al.)  Here we will only sketch
the theory of these branes, leaving a detailed
analysis for another occasion.

Given the WZW chiral algebra $\CA(G)$ it is natural to
choose a maximal torus $T\subset G$ and form the
parafermionic chiral algebra $\CA(G/T)$. This may
be defined by the coset construction with respect
to the level $\kbos$ torus chiral algebra $\CA(T)$.
The latter is obtained as follows.
We begin by  bosonizing the currents
in the maximal torus:
\eqn\pfi{
H^i(z) = -i \sqrt{\kbos} \p_z \phi^i  \qquad i = 1,\dots, r
}
where $r$ is the rank of $G$. The scalar $\phi$
has periodicity
\eqn\pfiv{
 \phi \sim \phi + 2 \pi \sqrt{\kbos} Q
}
where $Q$ is the root lattice of $G$. The full
chiral algebra $\CA(T)$ is extended by the
vertex operators
\eqn\pfii{
\exp[i \sqrt{\kbos} \alpha\cdot   \phi] \qquad \alpha\in Q
}
The representations of $\CA(T)$ are generated by
\eqn\pfiii{
\exp[i {1\over \sqrt{\kbos}} \mu\cdot   \phi ] \qquad \mu\in P
}
where $P$ is the weight lattice of $G$. The characters of these
representations   are level $\kbos$ theta functions
$\Theta_{\kbos,\mu}/\eta^r$. The label $\mu$ on representations
is subject to   $\mu\sim \mu + \kbos Q$. Thus the representations
are labelled by $P/\kbos Q$. For $SU(N)$ this has cardinality
$N\kbos^r$.

There is now a straightforward procedure for producing the
$\CA(G)/\CA(T):= \CA(G/T)$ coset model. The characters are
obtained from string functions defined by:
\eqn\pfv{
\chi_{\Lambda}(\tau,z) = \sum_{\mu\in P/\kbos Q} c_{\Lambda,\mu}(\tau)
{\Theta_{\kbos,\mu}(\tau,z)\over
\eta^r(\tau)}
}
For $G=SU(N)$, the parafermionic representations are subject to a
$Z_N$ selection rule. $\Lambda-\mu \in Q$, as well as   a  $Z_N$
identification from the outer automorphism of the Lie algebra
$(\Lambda, \mu) \sim (\Lambda' , \mu + \kbos \lambda_1)$ where
$\Lambda\to \Lambda'$ is defined in \automorph.

The finite group $\Gamma = Q/\kbos Q$ acts as a group of global
symmetries on both the $G/T$ theory  and the $T$ theory, with
$\beta\in Q/\kbos Q$ acting as
\eqn\globalsym{
\eqalign{
\Psi_{\Lambda,\mu} &
\to  e^{2\pi i \mu \cdot \beta/\kbos } \Psi_{\Lambda,\mu} \cr
\Psi_\mu & \to e^{-2\pi i \mu \cdot \beta/\kbos } \Psi_\mu\cr}
}
If we take the orbifold chiral algebra with this action we obtain
the original $G$ WZW chiral algebra, $(\CA(G/T) \otimes
\CA(T))/\Gamma \cong \CA(G)$. Therefore one can form branes in
the $G$ WZW theory using branes from the parafermionic theory.
The advantage of this viewpoint is that we can construct new
branes which do not have simple boundary conditions in terms of
the currents.\foot{It might be interesting to understand whether
these boundary fixed point theories are of use in the
multi-channel Kondo problem.}

For example, the
symmetry-preserving branes are of the form
\eqn\atype{
\vert \Lambda \rara_C^{G} =
 \sqrt{{1\over \vert \Gamma \vert}} \sum_{\mu\in P/\kbos Q} \vert
 \Lambda,\mu\rara_C \vert \mu\rara_C
}
However, now we can form the
 ``B-type branes'' in the language of \mmspara.
For example, one can now perform T-duality with respect to
the various $U(1)$ currents in the $\CA(T)$ theory.
For example, let $G=SU(N+1)$ and   consider the embedding
$SU(N) \times U(1) \hookrightarrow SU(N+1)$ where
$U(1)$ corresponds to the generator $H^N =
{1\over \sqrt{N(N+1)}} {\rm Diag}\{ 1,\dots, 1,-N\}$ of
the Lie algebra. Performing $T$-duality
with respect to the current $H^N(z)$ we get
\eqn\btype{ \vert \Lambda\rara_C^{',G}=
 \sqrt{{1\over\vert \Gamma \vert}} \sum_{\mu\in P/\kbos Q }
 \vert \Lambda,\mu\rara_C \vert \mu\rara_C' }
here $ \vert \mu\rara_C'$ is a state in the $\CA(T)$ theory. It
is $A$-type with respect to the first $(N-1)$ currents and
$B$-type with respect to $H^N(z)$, and only depends on $\mu -
(\alpha_N\cdot \mu)\lambda_N$.

To see that \btype\ is a qualitatively new brane we will
sketch the shape computation, showing that it can
be a $(2N+1)$-dimensional object in $G$.
The state \btype\ satisfies boundary conditions:
\eqn\newbdry{
\eqalign{
(J_n^a + \tilde J^a_{-n})\vert  B \rara & = 0 \qquad a\in su(N)\cr
(H^N_n - \tilde H^N_{-n}) \vert B \rara & = 0 \cr}
}
(but does not satisfy simple boundary conditions 
with respect to the remaining currents). 
Let us now work infinitesimally and consider the ``shape
computation'' for group elements $g = 1 + \sum_{j=1}^N (z^j
e_{N+1,j} - \bar z^j e_{j,N+1}) + \cdots $ where $z^j$ are
complex numbers and $e_{ij}$ are matrix units. We would like to
show that the overlap with the Cardy states $\langle g \vert \hat
\lambda, \hat \mu=0\rara^B_C$ is independent of $z^j$.  To show
this, note that the zeromode part of the ``B-branes'' necessarily
have $H_0^N \vert B \rara = \tilde H_0^N \vert B \rara =0$. Now,
in any $SU(N+1)$ representation $\Lambda$ if $H^N \vert \mu_L \ra
= H^N \vert \mu_R\ra =0$ then
\eqn\zeromtrx{ (\alpha^{(N)} + \alpha^{(j)})_N \langle \mu_L \vert
\rho_\Lambda(e_{j,N+1})\vert \mu_R \rangle = \langle \mu_L \vert
\rho_{\Lambda}([H^N, e_{j,N+1}] ) \vert \mu_R \rangle=0 }
This shows that $\langle g\vert \Lambda\rara_C^{',G}$ is
independent of $z^j$, to first order in $z^j$. It is important to
include higher order terms in $z^j$, but we will not do this here.

We conjecture that, at least for some $\Lambda$, the brane
described by \btype\  wraps a homologically nontrivial cycle in
$SU(N+1)$ transverse to the $SU(N)$ subgroup. This cycle is
explicitly the union of conjugacy classes:
\eqn\conjclss{
w_{2N+1} = \cup_{g\in SU(N+1),z\in U(1)} g {\rm Diag}\{ z^{-1/N},
\dots, z^{-1/N} , z^{1-1/N} \} g^{-1}
}
This is also the set of unitary matrices of the form $g_{ij} =
z^{-1/N}( \delta_{ij} + (z-1) v_i \bar v_j )$ where $v_i$ is a
unit vector in $C^{N+1}$ and $z\in U(1)$. It is the image in
$SU(N+1)$ of a  continuous map $S^1 \times CP^{N} \to SU(N+1)$
which factors through the unreduced suspension.

\newsec{Comparison with K-theory }

\subsec{The result of M. Hopkins}

Here we describe the result of a recent computation
of M. Hopkins. As we have mentioned, $SU(N)$ is
rationally a product of odd spheres $S^3 \times S^5
\times \cdots \times S^{2N-1}$. Branes cannot
wrap a homology cycle with $S^3$ as a factor, because
of the $H$-flux, but they can wrap all other cycles
(sometimes, we require finite multiplicity, as we have
seen).
This picture of D-branes
in $SU(N)$ at level $k$ suggests a charge group
\eqn\truecharg{
(Z/d_{k,N} Z)\otimes \Lambda_Z[w_5, \dots, w_{2N-1}]
}
More explicitly, the charge group is
\eqn\exterior{
 Z_{d_{k,N}} \oplus Z_{d_{k,N}} w_5
\oplus Z_{d_{k,N}} w_7 \oplus Z _{d_{k,N}}w_9 \oplus Z_{d_{k,N}}
w_{11} \oplus Z_{d_{k,N}} w_5 w_7 \oplus \cdots }
One can identify the $w_i$ with branes wrapping
the primitive homology cycles of
dimension $i$, generating the homology of $SU(N)$. The result \truecharg\ 
for the $K_H$-homology of $SU(N)$
has, in fact, been obtained recently by M. Hopkins \hopkins, using
a standard cell decomposition of $SU(N)$ \epstein, together with
the Meyer-Vietoris sequence. The cells are given by the
construction in \conjclss.

The computation of Hopkins leads to the formula
\eqn\hopenn{
d_{k,N}  = {\rm gcd}[{k\choose 1}, {k\choose 2},
\dots, {k \choose N-1}].
}
In appendix C  we show that this expression
coincides with
the expression found by Fredenhagen and Schomerus,
remembering that $\ksusy = \kbos + N$
\eqn\agree{
d_{k,N} = n(\kbos,SU(N))
}
In order
to compare with the K-theory formalism
we must use the {\it supersymmetric}
$SU(N)$ WZW model at the level $\ksusy $, defined by
the coupling to the $B$ field in the worldsheet Lagrangian
with supersymmetric fermions.

\subsec{Comparison for $SU(3)$}

The result of Hopkins confirms our computation
of $K_H^*(SU(3))$ in section 4.4.
 Moreover we have constructed
branes corresponding the different K-theory
classes in this case.

In particular,  using the results of
section 6, we  can give   a
picture for the branes contributing
to the classes in
$K^1_H(SU(3))$. These are branes wrapping the
twinned conjugacy classes $\CO^\omega$. As we noted
above, these are $S^2$ bundles over $M_5$.   Since
$SU(2)$ intersects $M_5$ transversally we
can interpret these branes as wrapping $M_5$, but
blown-up along the transverse  $SU(2)$ into a bundle
of $S^2$'s over   $M_5$.

\subsec{Comparison for $SU(N)$}

We have not generalized the computation of section 4 to
$SU(N)$, but this is in principle possible.

The symmetry-preserving branes described in section
5 account for the first term in \exterior.
Branes associated to the outer automorphism
of $SU(N)$,  wrapping  twinned conjugacy classes,
might carry other K-theory charges but clearly
are not numerous enough to  fill out the full set of
K-theory classes predicted by \truecharg.

 We
conjecture that applying T-dualities to the
parafermionic branes in $SU(N)$ will lead to
representatives of all the K-theory classes.

\newsec{Discussion}

A potentially fruitful   open problem is that of formulating a
definition of the   group of ``charges'' of (all) branes and
fluxes allowed in string theory and M-theory. This charge group
would help to distinguish the different homotopy classes of
collections of branes and fluxes. By ``homotopy classes'' we mean
we identify brane configurations related to one another by
continuous deformations of parameters, e.g., continuously
deforming the position of a D-brane, or the gauge fields on a
D-brane. etc.

The present paper offers one approach to this problem.
In broad terms  the group of components should be the group
of
{\it  homotopy classes of anomaly-free brane configurations modulo
equivalence relations imposed by dynamically allowed effects. }

This ``physical definition'' of a   group of components
cannot fail to be
right, but it is also not very precise or useful.
 We can make it more precise by focusing
 on D-branes at weak coupling. Then we have seen in this
paper that the charge group can be viewed as
a ``quotient'' of classes of anomaly-free configurations
of branes by identifications by D-instanton effects.
This philosophy offers, perhaps, an alternative route to
the definition of K-theory, or of the appropriate notion
which will replace K-theory when one is not working at
zero string coupling. It is possible that all the essential
physical effects determining D-brane charges have already been
discovered. If this is so, then this approach to a precise
mathematical framework for D-brane charges has a reasonable
chance of succeeding.

\vskip 4cm
 \centerline{\bf Acknowledgements}

GM  would like to thank I. Brunner, E. Diaconescu, D. Freed,
M. Hopkins, P. Landweber,  D. Morrison, G. Segal,
and E. Witten
for many important discussions on the topology
and K-theoretic interpretation of D-branes.
We thank D. Freed for comments on the draft. 

GM and JM  would like to thank the ITP at Santa Barbara for
hospitality during the writing of part of this manuscript.
This research was supported in part by the National Science
Foundation under Grant No. PHY99-07949 to the ITP
by  DOE grant DE-FG02-96ER40949 to Rutgers.

JM and NS are supported  in part by DOE grants
\#DE-FG02-90ER40542, \#DE-FGO2-91ER40654, NSF grant PHY-9513835
and the David and Lucile Packard Foundation.

\appendix{A}{Flux stabilization at higher rank}

In this appendix we describe a generalization of
the computation of \bds\ to higher rank groups.
For some steps we specialize to $G=SU(N)$, but
this restriction could probably easily be eliminated.

\subsec{Some volume computations}

In order to compare energies we will need volumes
of groups and homogeneous spaces.  We restrict
attention to $G=SU(N)$.

If we identify $T_eG $ with the Lie algebra
$\lieg$ of traceless, $N\times N$ antihermitian
matrices we will take the metric in this section
to be
\eqn\metrici{
g(X,Y) = - {\Tr}_N (XY)\qquad X,Y\in \lieg.
}
Put differently,   we choose the metric
\eqn\metric{
ds^2 = - Tr_N [(g^{-1} dg) \otimes (g^{-1} dg)] .
}

\bigskip
$\underline{\rm Volume\ of\ SU(N)  } $
\bigskip

We now compute the volume of  $SU(N)$ in the
metric \metric.  Consider $SU(N)/SU(N-1) = S^{2N-1}$.
We relate the local coordinates of $S^{2N-1}$ and $SU(N)$ via
\eqn\localcoord{
g= 1 + iy^N Diag\{\epsilon,\dots, \epsilon,1\} + \sum_i
(z^i e_{iN} - \bar z^i e_{Ni}) + \cdots
}
at $g=1$. (Here $\epsilon= -1/(N-1)$. ). Here we are thinking
of $S^{2N-1}$ as the solutions to $\sum_{i=1}^N
\vert z^i\vert^2=1$ and
$z^N = x^N + i y^N$.

In these coordinates, the group metric \metric\ becomes
\eqn\metrspher{
ds^2 = 2 \sum_{i=1}^{N-1} [(dx^i)^2 + (dy^i)^2] + {N\over N-1} (dy^N)^2
}

While \metrspher\ is not the round metric on the sphere it does
give the invariant volume form on the sphere, up to an overall
multiple. By left invariance we can compute this factor at the
identity coset.
Therefore, in the metric \metric\ we have
$$
\vol(SU(N)/SU(N-1)) = (2^{N-1} \sqrt{N\over N-1}) \vol(S^{2N-1}) =
 \sqrt{N\over N-1} {(2\pi)^N\over \Gamma(N)}
$$
(where we have embedded $SU(N-1)\hookrightarrow SU(N)$ with
index 1).

Therefore
$$
\vol(SU(N)) = {\sqrt{N}\over 2\pi} (2\pi)^{\half N(N+1)}
{1\over 1! 2! 3! \cdots (N-1)!}
$$

In particular,
$$
\eqalign{
\vol(SU(3)) &= \half \sqrt{3} (2\pi)^5\cr}
$$

This checks against a formula in the literature \yokota,
which was worked out in a slightly different way.

\bigskip
$\underline{\rm Volume\ of\ G/T } $
\bigskip

The map $G/T \times T \to G$ is (generically) an $N!$-fold
covering. We compute the volume of $T\cong U(1)^{N-1}$ by
taking standard orthonormal generators
$$
H_j = \sqrt{-1} {1\over \sqrt{j(j+1)} }Diag[1,\dots,
1, -j,0,\dots, 0]
$$
so the range of angles is $0\leq \theta_j\leq \sqrt{j(j+1)} 2\pi$.
Therefore
$$
\vol(T) = (N-1)! \sqrt{N}(2\pi)^{N-1}
$$
So $\vol(G/T) = N! \vol(G)/\vol(T)$ gives
\eqn\volgt{
\vol(G/T) = N (2\pi)^{\half N(N-1)} {1\over 1! 2! \cdots (N-1)!}
}

One can rewrite this in a way that holds for all Lie groups. The
positive roots are $e_i - e_j$ for $N\geq j> i\geq 1$. The
fundamental weights are
$$
\lambda_i = e_1 + \cdots + e_i -
{i\over N} \sum_{j=1}^N e_j
$$
The Weyl vector is:
$$
\rho = \sum_{j=1}^N \half(N- (2j-1)) e_j
$$
{}From this one computes
$$
\prod_{\alpha>0} \alpha\cdot \rho = 1! 2! \cdots (N-1)!
$$
so we can write
\eqn\volgtii{
\vol(G/T) = N \prod_{\alpha>0} {2\pi\over \alpha\cdot\rho}
}

One important subtlety is that  the above is the metric on
the space of left cosets $gT$. In our application we are
interested in the adjoint group acting on the
Lie algebra  $g X g^{-1}$. Since the center of the group acts
trivially we should compute the volume for $G=SU(N)/Z_N$.
This gives
\eqn\volgtiii{
\vol(G_{adj}/T) = \prod_{\alpha>0} {2\pi\over \alpha\cdot\rho}
}

\subsec{Flux stabilization at higher rank: Evaluation of the DBI
action}

We will work with $SU(N)$. The rank $r=N-1$. The
number of positive roots is $\Delta = \half N(N-1)$.

We now choose coordinates
\eqn\gpar{
g = k(\theta)^{-1} t(\chi)  k(\theta)
}
where $\chi$ are coordinates along the Cartan torus, and
$\theta$ are angular coordinates along the space $G/T$.
To be specific, we introduce a Cartan-Weyl basis
 \eqn\basislie{
 \eqalign{
 [H_i, H_j] & = 0 \cr
 [H_i, E_{\alpha}] & = \alpha^i E_{\alpha}\cr
 [E_\alpha , E_\beta] & = \cases{E_{\alpha + \beta} &$\alpha +
 \beta $ is a root\cr
                                     0& otherwise} \cr
 [E_{\alpha}, E_{\bar \beta}]& = \cases{\alpha^i H_i & $\alpha =
 \beta $\cr
 0 & otherwise } \cr} }
where the notation is that $\alpha, \beta,\dots $ label positive
roots and $\bar \alpha :=-\alpha$ etc. are the negative roots. We
are also normalizing the generators so that
\eqn\tracenorm{
\eqalign{
{\Tr}(H_i H_j) & = \delta_{ij} \cr
{\Tr}(E_\alpha E_{\bar \beta}) & = \delta_{\alpha, \beta} \cr}
}
Finally we set $t(\chi) :=\exp[i \chi_m H_m] $ which  are good
coordinates transverse to   the adjoint orbit $\CO_{\chi}$ of
$\exp[i\chi\cdot H]$ in $G$. Moreover we define
\eqn\expdholo{
 dk k^{-1}= \theta^{\alpha} E_\alpha  - \bar \theta^{\alpha}
 E_\alpha^\dagger + i \rho^i H_i . }
The 1-forms $\theta^\alpha$ span a basis of holomorphic $(1,0)$
forms on $\CO_{\chi}$ in the natural complex structure on $G/T$
induced by a choice of simple roots. Using $$ g^{-1} dg = k^{-1}
dk - (tk)^{-1} dk k^{-1} (tk) + k^{-1} (t^{-1}dt)k $$ and a
little algebra one can show that the  2-form
\eqn\formbee{ b = {1\over 4\pi^2 i}   \sum_{\alpha>0} \biggl(
\chi_m \alpha^m - \sin(\chi_m \alpha^m)\biggr) \theta^\alpha
\wedge \bar \theta^{\alpha} }
precisely reproduces
\eqn\cartanform{
x_3 = {1\over 24\pi^2} {\Tr}_N(g^{-1} dg)^3 = d b
}
in an open  neighborhood of the identity in a compact Lie group
$G$.

To get normalizations straight, note that
if $H=dB$ is an integral class and $F/(2\pi)$ is an
integral class then   the worldsheet path integral
looks like
$$
\exp[ i \int_{\Sigma} 2\pi B+ F ]
$$
So we should have
:
\eqn\effbee{ {F\over 2 \pi} +B = {k \over 2\pi} \sum
\sin(\chi\cdot \alpha) {\theta_\alpha \wedge \theta_{\bar
\alpha}\over 2\pi i} }
for $F = k \sum \chi\cdot \alpha
{\theta_\alpha \wedge \theta_{\bar \alpha}\over 2\pi i} $. Note
that flux quantization restricts $\chi = 2\pi (\lambda +
\rho)/k$, as expected.

In the parametrization \gpar\ the orbit
has metric
induced by \metric:
\eqn\metricorbit{ -{\Tr}_N(g^{-1} dg\otimes g^{-1}
dg)\vert_{\CO_\chi}= 4\sum_{\alpha>0} \sin^2 (\half \alpha\cdot
\chi) (\theta_\alpha \otimes \theta_{\bar \alpha} + \theta_{\bar
\alpha} \otimes \theta_{  \alpha}) }

As described below, the standard
 group metric \metric\  should be rescaled by $k/2$ so
\eqn\dbione{ g + 2\pi(F  +2\pi B ) = 2k \sum_{\alpha>0} \Biggl[
\sin^2 (\half \alpha\cdot \chi) (\theta_\alpha \otimes
\theta_{\bar \alpha} + \theta_{\bar \alpha} \otimes \theta_{
\alpha}) + i\half \sin \alpha\cdot\chi (\theta_\alpha \otimes
\theta_{\bar \alpha} - \theta_{\bar \alpha} \otimes \theta_{
\alpha}) \Biggr] }
and hence  we get:
\eqn\bdsextend{ d^{2\Delta} \xi \sqrt{\det \bigl[ g + 2\pi(F+2\pi
B)\bigr] } = \bigl({2k }\bigr)^{\Delta} (\prod_{\alpha>0} \sin
\half \alpha\cdot \chi) \vol(G_{adj}/T) }
where $\vol(G_{adj}/T)$ is the volume form for the
adjoint orbit $G_{adj}/T \subset \lieg$ in the metric \metric.

Now, we have shown that wrt the metric \metric\ the volume of
$G_{adj}/T$ is
\eqn\volgt{
 \prod_{\alpha>0} {2\pi \over \alpha\cdot \rho}
}
and so   we get:
\eqn\dbi{
 \int_{\CO_\chi} \sqrt{\det (g+F+B)}
=   \bigl({2k}\bigr)^{\Delta}
(\prod_{\alpha>0} \sin \half \alpha\cdot \chi)
\prod_{\alpha>0} {2\pi \over \alpha\cdot \rho}
 }

\bigskip
{\bf Remarks:}

\item{1.}
Notice that the $\sin$ factors come in the first power in \bdsextend.
This is to be contrasted with Weyl's formula for integrating class
functions:
\eqn\weyls{
\int_G f(g)dg = \int_T f(t) \prod_{\alpha>0} \sin^2(\half \alpha\cdot t)
{dt \over \vert W \vert}
}
which implies
that
\eqn\weyls{ \sqrt{\det g} \sim (\prod_{\alpha>0} \sin^2 \half
\alpha\cdot \chi){1\over N!} }
The flux-stabilization mechanism effectively takes a squareroot of the
volume!

\item{2.} The flux   on this orbit is
\eqn\bwb{
{F\over 2\pi} = \sum_{\alpha>0} (\lambda+\rho)\cdot \alpha
{\theta_\alpha \wedge \theta_{\bar \alpha}\over 2\pi i}
}
In the cohomology of $G/T$ this is $c_1(\CL_{\lambda+\rho})$ for
the line bundle $\CL_{\lambda+\rho} \to G/T$ one encounters in
the Borel-Weil-Bott theorem.

\subsec{Comparison with the CFT expression}

The normalization of the Minkowskian action for $SU(N)$ level $k$
WZW theory is
\eqn\wzwlag{ S = {k\over 8 \pi} \int_{\Sigma} dx dy  {\Tr}_N [
(g^{-1}\p_x g)^2 + (g^{-1}\p_y g)^2] + {2 \pi k } \int x_3 }

Since the action is ${1\over 4 \pi \alpha'} G_{\mu\nu} \p X^\mu
\p X^\nu$ and we are taking $\alpha'=1$ we have metric on the
group $SU(N)$ given by
\eqn\normmetric{
g = -{k \over 2} {\Tr}_N (g^{-1}dg\otimes g^{-1} dg)
}
for $SU(N)$.

Now we must multiply \dbi\ by the tension of the D-brane, which is
$1/(2\pi)^{2\Delta}$, (in units with   $\alpha'=1$). To avoid
confusions with string coupling and the normalization of the
Newton constant, we actually will just compute the ratio of the
mass of the brane to  the mass of a D0 brane with $\lambda=0$.
\foot{We have also left out the boundary states for the fermions.
These introduce extra factors of $\sqrt{2}$, but again cancel out
of the ratio.} The net result is
\eqn\energy{ {{\rm mass~ of~ D-brane~ on ~ } \CO_\chi \over {\rm
mass~ of~ D0-brane} } =
 k^{\Delta}
(\prod_{\alpha>0} \sin \half \alpha\cdot \chi)
\prod_{\alpha>0} {1 \over \pi \alpha\cdot \rho}
 }

Let us now compare this with the exact expression
from conformal field theory.
 For $SU(N)$ we have the formula:
\eqn\sunmtrx{ S_{0,\lambda} = {2^\Delta\over \sqrt{N}
(k+N)^{r/2}} \prod_{\alpha>0} \sin\biggl(\pi {\alpha\cdot
(\lambda+ \rho)\over k+N}\biggr) }
Note that $S_{00}\sim k^{-\half \dim G}$. Therefore,
the mass of the Cardy state $\vert \lambda \rara_C$ is
\eqn\crdymss{
{Energy(\vert \lambda\rara_C)\over Energy(\vert 0\rara_C)}=
{S_{0\lambda}\over S_{00} } = \prod_{\alpha>0}
{\sin\biggl(\pi {\alpha\cdot (\lambda+ \rho)\over k+N}\biggr)
\over \sin\biggl(\pi {\alpha\cdot  \rho\over k+N}\biggr)}
}
For $k\gg N $ we can expand the denominator in \crdymss\ to obtain
%
\eqn\appxmass{
(k+N)^\Delta \prod_{\alpha>0} {1\over \pi \alpha\cdot \rho}
\prod_{\alpha>0}
\sin\biggl(\pi {\alpha\cdot (\lambda+ \rho)\over k+N}\biggr)
}
Comparison with \energy\ shows that the  leading $k$-dependence
 fits perfectly with the interpretation of a brane wrapping
$G/T$ once.

\appendix{B}{Volume of $M_5$}

\subsec{Integrally-normalized left-invariant forms}

All traces in this section are in the $N$ of $SU(N)$.
We know that
\eqn\xthree{
x_3 = - {1\over 24\pi^2}{\Tr}(g^{-1} dg)^3
}
is the integral generator of $H^3(SU(N);Z)$. In general the
spherical class in $\pi_{2k+1}(SU(N))$ is $k!$ times the primitive
generator of $H_{2k+1}(SU(N))$ for $N$ in the stable range
\refs{\botti,\bottii,\smith}. Using this and the ABS construction
one can find the integral normalization of the higher traces. In
particular we have the integral class
\eqn\xfivenorm{
x_5 = -2 {2! \over (2\pi i)^3 5!}   {\Tr}(g^{-1} dg)^5
}
In the coordinates
\eqn\coords{
g= \exp[i \theta_1 H_1 + i \theta_2 H_2 + \sum_{\alpha>0}
(z_\alpha E_{\alpha} - \bar z_\alpha E_{\bar \alpha}) ]
}
 we find near the identity
\eqn\fifthpower{
\eqalign{
{\Tr}(g^{-1} dg)^5 & = {15 i \over \sqrt{2}} d\theta_1
( d^2 z_1 d^2 z_2 + d^2 z_1 d^2 z_3) \cr
& +
{15 i \over \sqrt{6}} d\theta_2 ( d^2 z_1 d^2 z_3 -2  d^2
z_2 d^2 z_3 -d^2 z_1 d^2 z_2) \cr
& -{15\over \sqrt{3}} d\theta_1 d\theta_2 (dz_1 dz_2
d\bar z_3 - d\bar z_1 d\bar z_2 dz_3)\cr
& + {\Tr} p^5 \cr}
}
 where $d^2 z := dz \wedge d\bar z$.
Here $p = dz_\alpha E_{\alpha} - d\bar z_\alpha E_{\bar \alpha}$.

Now, since $x_5$ restricts to the left-invariant
unit volume form on $M_5$ we can compare with that induced by the
metric \metric. From \fifthpower\ we get
$$
x_5\vert_{M_5} = - {4 \over (2\pi)^3 5!}{30 \over \sqrt{3}}
d\theta_1 d\theta_2 dy_1 dy_2 dy_3
$$
whereas the Haar measure in the metric \metric\ is
$\sqrt{8} d\theta_1 d\theta_2 dy_1 dy_2 dy_3$. From this
we deduce   that the induced volume on
$M_5 \subset SU(3)$ is
\eqn\volmfvnew{
\vol(M_5) = {4 \over \sqrt{2}}\sqrt{3}  (2\pi)^3
}

\appendix{C}{Proof of an arithmetic identity}

We would like to prove that
\eqn\fs{
{k\over
{\rm gcd} (k, {\rm lcm}(1,2,\dots, N)) }
}
is the same as
\eqn\hopkins{
{\rm gcd}[{k\choose 1}, {k\choose 2},
\dots, {k \choose N}]
}

Consider the prime divisors of $k$ and
of $1,2,\dots, N$. For each prime $p$ and
integer $n$ write
$n = p^{v_p(n)} n'$ where $n'$ is relatively
prime to $p$.
We will compare the prime powers in \fs\ and 
\hopkins\ and show that they are the same. 

We consider 3 cases:

1. If $v_p(k) = 0$ then both factors obviously
have $p^0$.

2. If $v_p(k)>0$ and $v_p(k) > v_p(j)$ for all
$j=1,2,\dots, N$ then note that
\eqn\identity{ {k\choose j} = {k\over j} {k-1\over 1} {k-2\over
2} \cdots {k-(j-1)\over j-1} = \pm {k \over j} (1-k)(1-k/2)\cdots
(1- {k\over j-1}) }
The first factor has $p^{v_p(k)-v_p(j)}$ while the others are
of the form $1+ p^{\rm positive}$, and the coefficients of $p^{\rm positive}$
are fractions whose denominators are prime to $p$. Therefore
$v_p({k\choose j}) = v_p(k)-v_p(j) $.

Now
$$
\eqalign{\min_{j=1,2,\dots, N} (v_p(k)-v_p(j)) &= v_p(k) -
\max_{j=1,\dots, N}(v_p(j))\cr
&= v_p(k) - \min[ v_p(k), \max_{j=1,\dots, N}(v_p(j))]}
$$
establishes the identity.

3. Finally, suppose that for some $j\leq N$, $v_p(j)\geq v_p(k)>0$.
Clearly, there is a $j'\leq j$ with $v_p(j') = v_p(k)$.
We claim that for this $j'$ we have
$v_p({k \choose j'}) = 0$. This again follows from \identity\ together
with the observation that every $j''<j'$ has $v_p(j'')<v_p(k)$.
Therefore, for such primes the power of $p$ in \fs\ and
\hopkins\ is $p^0$.

Thus \fs\ is equal to \hopkins\ $\spadesuit$

\listrefs

\bye